\documentclass[prd,twocolumn,amsmath,amssymb,floatfix,superscriptaddress,nofootinbib]{revtex4-1}

\usepackage{natbib}
\usepackage{amsmath}
\usepackage{amssymb}
\usepackage{latexsym}
\usepackage{bm}   
\usepackage{graphicx,epstopdf}
\usepackage{color}
\usepackage{hyperref}

\DeclareMathAlphabet\mathbfcal{OMS}{cmsy}{b}{n}
\definecolor{ultramarine}{rgb}{0.07, 0.04, 0.56}
\definecolor{cadmiumgreen}{rgb}{0.0, 0.42, 0.24}
\definecolor{indigo(dye)}{rgb}{0.0, 0.25, 0.42}
\hypersetup{
colorlinks=true,
citecolor=ultramarine,
linkcolor=cadmiumgreen,
urlcolor=indigo(dye)
}

\newcommand{\f}[2]{\frac{#1}{#2}}
\newcommand{\mk}[1]{\left( #1 \right)}

\newcommand{\be}{\begin{equation}}
\newcommand{\ee}{\end{equation}}
\def\U{{\cal U}}
\newcommand{\sV}{{\cal V}}
\newcommand{\curv}{{\cal R}}
\newcommand{\tq}{q}



\begin{document}

\preprint{}

\title{Running from Features: Optimized Evaluation of Inflationary Power Spectra}

\author{Hayato Motohashi}
\affiliation{Kavli Institute for Cosmological Physics, The University of Chicago, Chicago, Illinois 60637, USA}

\author{Wayne Hu}
\affiliation{Kavli Institute for Cosmological Physics, The University of Chicago, Chicago, Illinois 60637, USA}
\affiliation{Department of Astronomy and Astrophysics, University of Chicago, Chicago IL 60637, USA}

\begin{abstract}
In models like axion monodromy, temporal features during inflation which are not associated with its ending can produce
scalar, and to a lesser extent, tensor power spectra where deviations from scale-free power
law spectra can be as large as the deviations from scale invariance itself.  Here
the standard slow-roll approach breaks down since its parameters evolve on an
efolding scale $\Delta N$  much smaller than the efolds to the end of inflation.
Using the generalized slow-roll approach, we show that the expansion of observables
in a hierarchy of potential or Hubble evolution parameters comes from a Taylor expansion
of the features around an evaluation point that can be optimized.  Optimization of the leading-order
expression provides a sufficiently accurate approximation for current data 
as long as the power spectrum
can be described over the well-observed few efolds by the local tilt and running.   Standard second-order approaches, often used in the literature, ironically are worse than leading-order 
approaches due to inconsistent evaluation of observables.  We develop a new optimized next-order approach which predicts observables to $10^{-3}$ even for $\Delta N\sim 1$ where all parameters in the infinite hierarchy are of comparable magnitude.   For models with $\Delta N \ll 1$, the generalized slow-roll approach provides integral expressions that are accurate to second order in the deviation from scale invariance.  Their evaluation in the monodromy model provides highly accurate  explicit relations between the running oscillation amplitude, frequency and phase in the curvature spectrum and
parameters of the potential. 
\end{abstract}


\date{\today}

\maketitle

\section{Introduction}

In a general inflationary model, temporal scales that are not directly associated with the end of 
inflation leave their imprint in cosmological observables.  
For canonical single field models, these arise from features in the potential which
leave the curvature and gravitational wave power spectra nearly scale invariant but no longer 
scale-free power laws.    
Current observations constrain deviations from scale-free power law spectra only as
well as deviations from scale invariance itself
\cite{Ade:2015lrj}.   Therefore these constraints only impact models with large features.

In the standard slow-roll approximation, the leading-order effect would be a running of the tilt 
of which the amplitude is quadratic in the small deviation from scale invariance.  It is common
to test for a finite value of such a parameter in data sets like the Planck CMB power spectra
\cite{Ade:2015lrj}.   Yet the running of the tilt is only constrained at the $\sim 10^{-2}$ level,
the same level as the tilt from scale invariance itself.  Interpreting these constraints requires going beyond the standard  approximation where slow-roll parameters are taken to be both nearly constant
and strongly hierarchical (e.g.~\cite{Stewart:1993bc,Lidsey:1995np,Leach:2002ar,Cortes:2007ak}).  Using the standard second-order approach on such data, as is common in
the literature (e.g.~\cite{Planck:2013jfk,Ade:2015lrj,Baumann:2014cja,Gao:2014yra,Boyle:2014kba}),  can provide
misleading results for models where the running of the tilt is not exactly constant or the 
potential purely cubic.

On the other hand, if the observed
power spectra are well characterized by the local tilt and running of the tilt, which are the first two terms in 
a Taylor expansion of a continuous feature, then a slow-roll hierarchy still exists and
can be used to predict observables accurately (cf.~\cite{Dodelson:2001sh}).   
It is simply that the efolding scale of
the feature $\Delta N$, while still  greater than unity, is much less than the number of 
efolds to the end of inflation.

In this paper, we use
the generalized slow-roll (GSR) approach  \cite{Stewart:2001cd,Dodelson:2001sh,Gong:2001he,Choe2004,Dvorkin:2009ne,Gong:2004kd,Hu:2014hoa} to extend the validity of the standard slow-roll approximation.   Here the deviations from scale invariance are only assumed to be small in amplitude not in temporal frequency.    With GSR, we explore the 
relationship between features in the potential and the hierarchy of slow-roll parameters.
Based on a temporal Taylor expansion, we 
develop optimized approaches for the evaluation of curvature
and gravitational wave power spectra from inflation.    We test these approaches in the
axion monodromy model
\cite{Silverstein:2008sg} where
low frequency cases  produce a slowly varying running of the tilt 
\cite{Kobayashi:2010pz,Czerny:2014wua,Meerburg:2014bpa} and high frequency cases
imprint oscillations in the power spectra \cite{Flauger:2009ab,Flauger:2014ana}.

The paper is organized as follows.
In \S \ref{sec:GSR}, we review the GSR approach and iterate it to second order in the
amplitude of deviations from scale invariance to describe power spectra in terms of integrals
over the temporal history of inflation.
We consider first order terms in 
\S \ref{sec:first} and show how the Taylor expansion of the temporal history is related to observables and to 
the hierarchy of slow-roll parameters of the potential and Hubble evolution.   Optimized
evaluation of the leading-order Taylor term \cite{Stewart:2001cd} provides predictions and inflation potential reconstructions that are accurate to
next-to-leading order.  These suffice for models with running of the tilt that is nearly constant
across the observable scales.  
In \S \ref{sec:second}, we generalize the optimized approach to second order and an optimized 
next-to-leading-order approximation and contrast these approaches with the
standard second-order approximation which inconsistently treats these evaluations.
We illustrate these optimized evaluation techniques for monodromy in both the low
and high frequency regime in \S \ref{sec:monodromy}.  We discuss these results in
\S \ref{sec:discussion} and provide relationships between the various parameterizations 
in the Appendix.

\section{Generalized Slow Roll}
\label{sec:GSR}

The GSR approach is ideal for studying and extending the validity of the standard slow-roll approximation as it assumes only that the amplitude of deviations from a de Sitter expansion are small with no assumptions on their temporal frequency or equivalently on the smoothness of the potential \cite{Stewart:2001cd,Dodelson:2001sh,Choe2004,Dvorkin:2009ne}.  In this approach, we iteratively solve the exact Mukhanov-Sasaki
equation for the evolution of the inflaton mode function or field fluctuation
in the spatially flat gauge $\delta \phi= y/a\sqrt{2 k} $
\begin{equation}
{d^2 y \over dx^2} + \left( 1 - {2 \over x^2} \right) y = \left( \frac{f''- 3f'}{f} \right)
\frac{y}{x^2} ,
\label{eqn:yeqn}
\end{equation}
to obtain the comoving curvature power spectrum
\begin{equation}
\Delta_\curv^2 = \lim_{x \rightarrow 0} \left|\frac{ x  y}{f} \right|^2.
\end{equation}
Here $' = d/d\ln\eta$, $x = k\eta$ and $\eta$ is the (positive, decreasing) conformal time to the end of inflation and the deviations from de Sitter expansion are given by variations in
\begin{align}
\label{eqn:fdef}
f^2  & =  \frac{8 \pi^2 \epsilon_H}{H^2} \left( {a H \eta} \right)^2,
\end{align}
where the Hubble slow-roll parameter is defined by
\begin{equation}
\epsilon_H = - \frac{d\ln H}{dN} = \frac{1}{2}\left( \frac{d \phi}{dN} \right)^2,
\label{eqn:eps}
\end{equation}
with $N$ as the (negative, increasing) number of efolds relative to the end of inflation.

The fundamental assumption in this approach is that the mode function $y$ remains close
to its  de Sitter form 
\begin{equation}
y_0(x) = \left( 1 + {i \over x} \right) e^{i x} ,
\label{eqn:BDcond}
\end{equation}
and we will loosely refer to this property as requiring the amplitude of deviations from de Sitter
or scale invariance encoded in $f$ to be small (see \cite{Kinney:2005vj,Namjoo:2012aa,Martin:2012pe,Motohashi:2014ppa} for exceptional cases).
If so, we can 
take the formal Green function solution to Eq.~(\ref{eqn:yeqn}),
\begin{equation}
y(x) = y_0(x) - \int^\infty_x \f{du}{u^2} \frac{f''- 3f'}{f} y(u) {\rm Im}[y^*_0(u) y_0(x)]  ,
\end{equation}
replace $y \rightarrow y_0$ on the right-hand side
and iteratively improve the solution.

To first order in the de Sitter deviations, this procedure results in \cite{Kadota:2005hv}
\begin{equation}
\ln \Delta_\curv^2(k)
 \approx I_0 = G(\ln x_{*}) + \int_{x_{*}}^\infty {d x \over x} W(x) G'(\ln x),
 \label{eqn:GSR}
\end{equation}
where $x_* \ll 1$, 
\begin{equation}
G = - 2 \ln f    + {2 \over 3} (\ln f )'  ,
\label{eqn:G}
\end{equation}
and 
\begin{equation}
W(x) = {3 \sin(2 x) \over 2 x^3} - {3 \cos (2 x) \over x^2} - {3 \sin(2 x)\over 2 x} .
\end{equation}
While we assume a canonical scalar field throughout, all of the GSR-based results here and
below
can be readily generalized to $P(X,\phi)$ theories and the effective field theory that
parameterizes them \cite{Hu2011,Hu:2014hoa}.

The function $W$  determines exactly
how inflaton fluctuations freeze into curvature fluctuations as they pass the horizon.
We can make the connection more explicit by integrating Eq.~(\ref{eqn:GSR}) by parts to obtain
\cite{Kadota:2005hv}
\begin{equation}
I_0  \approx -\int_0^\infty \frac{d x}{x} W'(x) G(\ln x).
\label{eqn:GSRGint}
\end{equation}

Note that $G$ is a function of  time $\ln\eta$ alone and denoting it as a function of
$\ln x$ is simply a convenient choice of its zero point for a given $k$. 
Scale dependence of the power spectrum arises only if $G'\ne 0$ and hence its value
quantifies the deviations from de Sitter results.   These typically must be 
\begin{equation}
\bar G' = {\cal O} \mk{ \f{1}{N} }
\end{equation}
in order for features not to prematurely end inflation.   The overbar here represents
an average over several efolds and denotes
the fact that only
the  integral of $G'$ needs to be small or equivalently that transient effects do not
necessarily end inflation.   This average as we shall 
see is closely related to the average tilt of the curvature spectrum.
In fact Eq.~(\ref{eqn:GSRGint})  is valid even if the inflaton potential  contains discontinuities or
delta function sources in $G'$ where the local tilt and average tilt differ substantially.        Given that 
$\lim_{x\rightarrow0} W'=0$, this form also conserves curvature fluctuations outside
the horizon and gives a manifestly positive definite power spectrum, and it remains a controlled approximation for up to order
unity deviations  \cite{Dvorkin:2009ne} unlike related variants \cite{Stewart:2001cd,Choe2004}.

In particular, linearity in $\bar G'$ should not be conflated with the constancy of $G'$.  
If we define $\Delta N$ as the typical efolding scale of its variation, then
\begin{equation}
{G''} = {\cal O}\left(\frac{1}{\Delta N}\right) {G'} = {\cal O}\left(\frac{1}{N\Delta N}\right)
\end{equation}
rather than $1/N^2$ as the standard slow-roll approximation assumes by requiring that
the only temporal feature during inflation be associated with its end.  We shall show in \S \ref{sec:first} that if  $1 <  \Delta N \ll |N|$, a slow-roll hierarchy of successively smaller derivative parameters still exists, but order
counting needs to be generalized using a consistent Taylor expansion of $G$ between observables.   If $\Delta N \ll 1$, the hierarchy is inverted so that computing 
observables requires resumming an infinite series or equivalently  the
direct integration of  Eq.~(\ref{eqn:GSRGint}).  In this regime observables depend on
the continuous function $G$ (see \S \ref{sec:highfrequency}), and model-independent
approaches seek to reconstruct that function rather than measure a series of parameters
\cite{Dvorkin:2010dn,Dvorkin:2011ui,Miranda:2014fwa}.

These ${\cal O}(1/N\Delta N)$ effects should be distinguished from the 
 true $1/N^2$ ones.   In the GSR approximation, these come from terms that are quadratic in
 the deviation from de Sitter solutions $G'$.   Iterating the Green function approach, we 
 obtain \cite{Choe2004,Dvorkin:2009ne}
\begin{align}
\Delta^2_\curv & \approx e^{I_0}\left[ \left(
1 + \frac{1}{4}I_1^2 + \frac{1}{2} I_2 \right)^2 + \frac{1}{2} I_1^2 \right],
\label{eqn:GSRpower}
\end{align}
where the first-order term $I_0$ is defined in Eq.~(\ref{eqn:GSR}) and
 the second-order corrections are 
\begin{eqnarray}
I_1 &=& \f{1}{\sqrt{2}} \int_0^\infty \f{d x}{x} G'(\ln x) X(x) , \nonumber\\
I_2 &=& -4 \int_0^\infty \f{dx}{x} \mk{ X + \f{1}{3} X' } \f{f'}{f} \int_x^\infty \f{du}{u^2} \f{f'}{f} ,
\label{eqn:I1I2}
\end{eqnarray}
with 
\begin{equation}
X(x) = {3 \over x^3} (\sin x- x \cos x)^2 .
\end{equation}
In the following sections we keep order counting in $1/N$ and $1/\Delta N$ distinct.

For tensor fluctuations, the same GSR approach holds with the replacement of $f$
with
\begin{equation}
f_h^2 = \frac{ 2\pi^2}{H^2} (a H\eta)^2
\label{eqn:fh}
\end{equation}
in the construction of the source $G_h$ of the tensor fluctuations \cite{Gong:2004kd,Hu:2014hoa}.   With these substitutions, Eq.~(\ref{eqn:GSRpower})
then provides the tensor power spectrum in each polarization state
$\ln \Delta_{+,\times}^2(k)$.

\section{Optimized Leading- and First-Order Approximation}
\label{sec:first}

Utilizing and extending the techniques of Ref.~\cite{Stewart:2001cd},
we elucidate  the conditions under which
a slow-roll hierarchy of parameters for scalar and tensor power spectra observables 
exists and the relative size
of terms in the series in  \S \ref{sec:GSRleading}.
In \S \ref{sec:optimizednext}, we
show how the next-to-leading-order terms in the hierarchical expansion are generally large
whenever running of the tilt is comparable to the deviations in the tilt itself but can be absorbed
into an optimization in the time of
fluctuation freeze-out consistently between observables. 
The resulting  optimized leading-order description suffices for 
scalar and tensor spectra that can be described by a nearly constant running of the tilt even 
when it is of order the deviations from scale invariance represented by the tilt itself (cf.~\cite{Dodelson:2001sh}).   We relate this hierarchy to the Hubble slow
roll parameters in \S \ref{sec:hubblesr} and to the potential slow-roll parameters in
\S \ref{sec:potential}.

\subsection{Smoothness Hierarchy}
\label{sec:GSRleading}

The hierarchical structure of the slow-roll parameters stems from a smoothness assumption
for the deviations from a de Sitter expansion.    
For any smooth source of deviations $G(\ln x)$ in the GSR formalism, we can Taylor expand its form around
an epoch near horizon crossing $\ln x_f$ and integrate its effect
term by term.  The standard slow-roll approximation proceeds by assuming $\ln x_f=0$
but we shall see there are advantages to tuning this evaluation point to make it correspond
better to the freeze-out epoch.  Note that
a shift in this point corresponds to a shift in efolds of $\delta N \approx -\delta \ln x_f$.

For the first-order $\ln \Delta_\curv^2 \approx I_0$ term of  Eq.~(\ref{eqn:GSRGint}), this leads to 
\begin{equation}
\ln \Delta_\curv^2 
 \approx G(\ln x_f)  + \sum_{p=1}^{\infty}\tq_p(\ln x_f)  G^{(p)}(\ln x_f),
\label{eqn:Gseries}
\end{equation}
where $G^{(p)}$ denotes the $p$th derivative of $G$ with respect to $\ln x$ and 
\begin{equation}
\tq_p(\ln x_f) = -\frac{1}{p!} \int_0^\infty  \frac{d x}{x} W'(x) \left(\ln \frac{x}{x_f}\right)^p.
\end{equation}
These coefficients can be calculated using the generating function \cite{Stewart:2001cd}
\begin{align}
F(z,x_f) =& -\int_0^\infty \f{dx}{x} W'(x) \left( \frac{x}{x_f} \right)^z
\nonumber\\
& = (2 x_f)^{-z} \cos \mk{\f{\pi z}{2}} \f{3\Gamma(2+z)}{(1-z)(3-z)},
\label{eqn:intf}
\end{align}
so that
\begin{equation}
\tq_p(\ln x_f)= \frac{1}{p!} \lim_{z \rightarrow 0} \frac{ \partial^p F(z,x_f)}{\partial z^p}.
\end{equation}
To clarify the dependence of these coefficients on $\ln x_f$,
it is also useful to express $\tq_p(\ln x_f)$ in terms of the first term
\be \tq_1(\ln x_f) =\ln x_1 - \ln x_f, 
\label{eqn:q1}
\ee
where
\begin{equation}
\ln x_1 = \f{7}{3}-\ln 2-\gamma_E.
\end{equation}
Here $\gamma_E$ is  the
Euler-Mascheroni constant.
Since
\be \f{d\tq_p}{d\tq_1} = - \f{d\tq_p}{d\ln x_f} = \tq_{p-1}, \label{eqn:dq} \ee
$\tq_p$ is a $p$th degree polynomial of $\tq_1$.    Each degree introduces an 
extra constant of integration that is independent of $\ln x_f$.  It is convenient to define
these in terms of $\tq_p(\ln x_1)$ since $\tq_1(\ln x_1)=0$.  
The higher coefficients become
\be \tq_p(\ln x_f)=\frac{\tq_1^p(\ln x_f)}{p!} + \sum_{n=0}^{p-2} \f{\tq_{p-n}(\ln x_1)}{n!}\tq_1^n(\ln x_f). \ee
The next two terms are thus given explicitly by 
\begin{align}
\tq_2(\ln x_f) &= \frac{\tq_1^2(\ln x_f)}{2} + \frac{4 -3 \pi^2}{72} ,
\label{eqn:q2q3}\\
\tq_3(\ln x_f) &= \frac{\tq_1^3(\ln x_f)}{6} + \frac{4 -3 \pi^2}{72}\tq_1(\ln x_f) +\frac{55}{81} -\frac{\zeta(3)}{3} . \nonumber
\end{align}
We  optimize the slow-roll approximation below by choosing
$\ln x_f$ to set certain coefficients to zero.   
Equation~\eqref{eqn:dq} implies that setting $\tq_{p}(\ln x_f)=0$ makes $\tq_{p+1}(\ln x_f)$ take 
on extremal values as a function of 
$\ln x_f$.   Certain optimizations also have the benefit that the extremum is a minimum of 
$|\tq_{p+1}|$ and hence suppress the next correction.

Since
\begin{equation}
\frac{dG^{(p)}(\ln x_f)}{d\ln k} = -G^{(p+1)}(\ln x_f),
\end{equation}
the tilt and the running of the tilt are given by
\begin{eqnarray}
n_s-1 &\equiv& \frac{d\ln \Delta^2_\curv}{d\ln k} \approx -G'(\ln x_f) -  \sum_{p=1}^{\infty}\tq_p  G^{(p+1)}(\ln x_f) ,
 \nonumber\\
\alpha &\equiv& \f{dn_s}{d\ln k} \approx G''(\ln x_f) +  \sum_{p=1}^{\infty}\tq_p  G^{(p+2)}(\ln x_f) ,
\label{eqn:tiltrun}
\end{eqnarray}  
with the obvious continuation to the running of each successive quantity.  Since the 
$\tq_p$ coefficients are fixed given a choice of $\ln x_f$ for all $\ln k$, 
we  omit the $\ln x_f$ argument of $\tq_p$ for clarity and where no confusion will 
arise we also do so below for compactness.
A nearly constant tilt in $\ln k$ requires $|G''| \ll |G'|$, a nearly constant running
$|G'''| \ll |G''|$, and each gains its next-to-leading-order correction from the $\tq_1$ term.

For tensor fluctuations, we similarly have \cite{Gong:2004kd,Hu:2014hoa}
\begin{eqnarray}
\ln \Delta_{+,\times}^2(k) &\approx& -\int_0^\infty \frac{d x}{x} W'(x) G_h(\ln x) \nonumber\\
&\approx& G_h(\ln x_f) + \sum_{p=1}^{\infty} \tq_p G_h^{(p)}(\ln x_f),
\label{eqn:GSRGhint}
\end{eqnarray}
which likewise determines the tensor tilt and running of the tilt
\begin{eqnarray}
n_t \equiv \frac{d\ln \Delta_{+,\times}^2}{d\ln k} 
 &\approx &- G_h'(\ln x_f) - \sum_{p=1}^\infty \tq_p G_h^{(p+1)}(\ln x_f) , \nonumber\\
\alpha_t
\equiv \frac{d n_t}{d\ln k}
 &\approx & G''_h(\ln x_f) + \sum_{p=1}^\infty \tq_p G_h^{(p+2)}(\ln x_f).
\label{eqn:GSRt}
\end{eqnarray}

For a sufficiently smooth $G(\ln x)$ and $G_h(\ln x)$, the series expansion of the power spectra will rapidly converge. Since
\begin{equation}
\lim_{p \rightarrow \infty} \frac{\tq_p}{\tq_{p-1}} = -\frac{1}{2} ,
\label{eqn:qp}
\end{equation}
the criteria for convergence for the scalar power spectrum is
\begin{equation}
\lim_{p \rightarrow \infty} \left| \frac{G^{(p)}}{G^{(p-1)}} \right| < 2,
\label{eqn:convergence}
\end{equation}
and is similar for the tensor spectrum and tensor source $G_h$.  
The series is dominated by the first term for sufficiently small values for this
ratio, with each successive term suppressed by the smoothness scale
$\Delta N$, $|G^{(p)}/G^{(p-1)}| = {\cal O}(\Delta N^{-1})$. 
Therefore, the Taylor expanded form above is applicable for features  with
$\Delta N \gtrsim 1$. For high frequency features where $\Delta N < 1/2$ and the
convergence criteria are violated, the GSR integral formula must be developed on a case by case basis  (see \S \ref{sec:highfrequency} for the monodromy example).

\begin{table}[t]
\begin{tabular}{|c|c|c|c|c|} \hline  
& LO/SO &       OLO & ONO  & GSR\\ \hline
$p$ & $0$ & $1$ & $2$ & $\infty$ \\ 
$\Delta N$ & $\gtrsim 50$ & $\gtrsim$ few & $\gtrsim 1$ & all\\
$\ln x_p$ & $0$ & $ 1.06$ & $ 0.22$ & -- \\\hline
$\tq_1$ & $1.06$ & $0$ & $0.84$  & --\\
$\tq_2$ & $0.21$ & $\ -0.36\ $ & $0$ & --\\ 
$\tq_3$ & $0.10$ & $0.28$ & $\ 0.078\ $ &-- \\ \hline
\end{tabular}
\caption{Slow-roll approximations are characterized by their order $p$ in a temporal
Taylor expansion which determines their applicability in describing features spanning $\Delta N$ efolds.  Standard second-order (SO) approaches only improve on leading order (LO) 
by keeping $p>0$ in some but not all observables.
Optimized evaluation (OLO/ONO) achieves consistent $p$th order accuracy with only $p-1$ additional parameters by evaluating them at special values of $\ln x_f=\ln x_p$.  The  coefficients $\tq_{n}$ for
$n>p$ control the error from truncation.  
The GSR approximation forgoes the Taylor expansion for an integral approach.
}  
\label{tab:optimized} 
\end{table} 

\subsection{Optimized Leading vs.\ Next Order}
\label{sec:optimizednext}

We call an approximation scheme that just retains the GSR  $I_0$ term and
no corrections from the Taylor series, first order in deviations from scale invariance and leading order (LO) in the hierarchy.  Namely,
\begin{eqnarray}
\ln \Delta_\curv^2 & \approx & G(\ln x_f),  \nonumber\\
n_s-1 &\approx & -G'(\ln x_f) ,\nonumber\\
\alpha &\approx & G''(\ln x_f) , \qquad {\rm (LO)}
\label{eqn:LOG}
\end{eqnarray}
and similarly for the tensor observables.
One that adds $p=1$ terms we call next(-to-leading)-order (NO)  in the
hierarchy.
These distinctions still allow us to choose the
evaluation epoch $\ln x_f$ which can be exploited to make a specific LO approximation as
accurate as the generic NO approximation.

In the standard slow-roll approximation, one takes $\ln x_f=\ln x_0\equiv 0$
and to NO 
\begin{eqnarray}
\ln \Delta_\curv^2 & \approx & G(0) + \tq_1(0) G'(0), \nonumber\\
n_s-1 &\approx & -G'(0) - \tq_1(0) G''(0),\nonumber\\
\alpha &\approx & G''(0) + \tq_1(0) G'''(0) .
\label{eqn:optimizedsr}
\end{eqnarray}
Here we have restored the $\ln x_f$ argument of
$\tq_n$ for clarity.
Since $\tq_1(0) \approx 1.06$ (see Table~\ref{tab:optimized}),
if $\alpha$ is comparable to $n_s-1$ then the NO corrections are comparable
to LO and are therefore required for accuracy.

On the other hand, since all observables follow this same Taylor
series form, this NO approximation is equivalent to shifting the
evaluation epoch by $q_1(0)$.   This brings the evaluation point to
\begin{equation}
\ln x_1 = \tq_1(0)  \approx 1.06, \qquad ({\rm OLO})
\end{equation}
which using Eq.~(\ref{eqn:q1}) is equivalent to setting 
\begin{equation}
\tq_1(\ln x_f) =\tq_1(\ln x_1)=0.
\end{equation}
We call this the optimized leading order (OLO) approximation.  
By adopting this optimization,
we gain all of the benefits of a next order approximation without any of the complexity.
From this point forward,
unless we specify otherwise, the LO approximation will refer to the {\it standard} LO approximation
of $\ln x_f=0$.

Beyond the NO
approximation of Eq.~(\ref{eqn:optimizedsr}), this shift does not resum the terms in the Taylor series since for $p>1$
\begin{equation}
\tq_p(0)\ne \frac{\tq_1^p(0)}{p!} 
\end{equation} 
or equivalently for the OLO evaluation point $\tq_{p}(\ln x_1) \ne 0$.
Furthermore $|\tq_2(\ln x_1)|$ is a local maximum of $|\tq_2(\ln x_f)|$ so that zeroing
the NO correction comes at the expense of maximizing the next-to-NO correction.
For a low frequency $\Delta N\gg 1$, i.e.\ for a strong hierarchy $|G^{(p)}/G^{(p-1)}|\ll 1$,
this is a small price to pay.
For $\Delta N \sim 1$, high accuracy requires going beyond the OLO optimization.
In \S \ref{sec:second} we will optimize the NO approximation itself by zeroing
the next-to-NO correction from $\tq_2$.   We shall see that this has the added benefit
that one can choose the solution that is a local minimum of $|\tq_3|$.

It is important to note that this series of optimizations is not equivalent to truncating at a fixed
order in 1/$\Delta N$.   Since
\begin{equation}
G^{(p)} \equiv \frac{d^p G}{d\ln x^p} = {\cal O}\left( \frac{1}{N \Delta N^{p-1}}\right),
\end{equation}
dropping ${\cal O}(1/N\Delta N)$ would eliminate $G''$ in all observables
including $\alpha$ and dropping ${\cal O}(1/N\Delta N^2)$ would retain the $\tq_1$ correction
for $n_s-1$ but not $\alpha$.
By truncating at the same order in the Taylor expansion of each observable,
we have ensured that they are all consistently evaluated at the same effective epoch
when the corrections are resummed.
We shall see that this feature is crucial for ensuring consistency between observables.
Table~\ref{tab:optimized} summarizes the various approximations and their applicability.

Since derivatives with respect to $\ln \eta$ and $\ln k$ are interchangeable, the
accuracy of this truncation at NO or equivalently OLO
is directly related to the accuracy with which 
the power spectrum can be described by the local tilt and running around 
a given pivot scale $k_0$ 
\begin{eqnarray}
\ln \Delta_\curv^2(k) & \approx & \ln A_s + [n_s(k_0)-1]\ln \left( \frac{k}{k_0}\right) \nonumber\\
&& + \frac{\alpha(k_0)}{2} \ln^2\left(\frac{k}{k_0}\right).
\label{eqn:local}
\end{eqnarray} 
If the observed
power spectrum is a good fit to this form over at least the $\Delta N\sim$ few that are observationally  well constrained, then OLO will also provide a good approximation.
Omitted corrections are suppressed relative to OLO by $1/\Delta N^2$ for each observable.\footnote{The apparent contradiction with the  power law example in Ref.~\cite{Dodelson:2001sh} is due to numerical problems in their calculation [E. Stewart (private communication)].  Their feature model calculation is also in error.} 
We shall quantify this consideration in \S \ref{sec:lowfrequency}.

\subsection{Hubble Slow Roll Parameters}
\label{sec:hubblesr}

Although the $G^{(p)}(\ln x_f)$ terms can themselves be thought of as the hierarchy of slow-roll parameters that are the most directly related to observables, it is useful to reexpress them in terms of the  more familiar Hubble slow-roll parameters $\epsilon_H$ and
\begin{equation}
\delta_p \equiv \frac{1}{H^p \dot\phi} \left( \frac{d}{dt} \right)^{p+1} \phi,
\end{equation}
or equivalently derivatives of $\ln H$ with respect to efolds 
through the hierarchy relations
\begin{eqnarray}
\frac{ d \ln \epsilon_H}{d N } & = &  2(\epsilon_H + \delta_1)  ,\nonumber\\
\frac{ d \ln  \delta_p}{dN } &=& \frac{\delta_{p+1}}{\delta_p}  + p \epsilon_H - \delta_1 .
\label{eqn:hubbleeom}
\end{eqnarray}
Using these relations we can express $G^{(p)}$ in terms of the Hubble
slow-roll parameters $\epsilon_H,\delta_p$ as detailed in the Appendix.  
In particular, the hierarchy of $G^{(p)}$ derivatives is equivalent to the hierarchy of
$\delta_p$ parameters.  Equation~(\ref{eqn:hubbleeom}) says that a hierarchical scaling with $|\delta_{p+1}/\delta_p| <1$  requires 
the fractional change in $\delta_p$ per efold to be small in addition to $\epsilon_H, |\delta_1|\ll 1$.   In particular, the Taylor expansion in $G^{(p)}$ becomes a summation over
the hierarchy $\delta_p$ (see the Appendix)
\begin{equation}
G^{(p)} \approx  (-1)^p \left( 2 \delta_p + \frac{2}{3} \delta_{p+1} \right) + {\cal O}\mk{\f{1}{N^2}},
\end{equation}
for $p\ge 2$. 
For order
counting purposes
\begin{equation}
\epsilon_H, \delta_1 = {\cal O}\left( \frac{1}{N} \right), \quad
\delta_p =  {\cal O}\left( \frac{1}{N\Delta N^{p-1}} \right).
\end{equation}

Although our expressions for observables
in terms of $G$ are explicitly ${\cal O}( G') = {\cal O}(N^{-1})$ due to the linearity of
Eq.~(\ref{eqn:Gseries}),
we can choose whether or not to keep nonlinear terms in the conversion of Hubble
slow-roll parameters to $G$.    Different choices will differ by ${\cal O}(N^{-2})$ and 
hence include a subset of second-order  corrections arising from the evaluation of the first-order GSR term.
Our convention will be to convert the OLO expression to ${\cal O}(N^{-1})$ in 
$\ln \Delta_\curv^2$ and $n_s-1$ but retain ${\cal O}(N^{-2})$ terms in $\alpha$.   This
is to maintain backward compatibility with the standard slow-roll approximation 
where leading order for $\alpha$
is assumed to be ${\cal O}(N^{-2})$.   
We also drop terms that are
${\cal O}(N^{-2}/\Delta N)$.  Thus, our leading-order expression spans the range of
possibilities from $\Delta N \ll |N|$ to $\Delta N \sim |N|$, i.e.~large to small running of the tilt.
In this sense, our OLO approach is more general than the optimization introduced in
Ref.~\cite{Stewart:2001cd} where $\Delta N \ll |N|$ is assumed.

With this convention the OLO scalar observables are
\begin{eqnarray}
\ln \Delta_\curv^2 &\approx&  \ln \frac{H^2}{8\pi^2\epsilon_H} -  \frac{10}{3} \epsilon_H  - \frac{2}{3} \delta_1 \Big|_{x= x_1} ,\\
\label{eqn:OLOHubble}
n_s-1 &\approx &-4\epsilon_H -2\delta_1 - \frac{2}{3} \delta_2  \Big|_{x= x_1} , \qquad ({\rm OLO}) \nonumber\\
\alpha & \approx & 
 -2 \delta_2 -\frac{2}{3} \delta_3  -8 \epsilon_H^2  -10\epsilon_H\delta_1 + 2\delta_1^2  \Big|_{x= x_1}. \nonumber
\end{eqnarray}
Here and below the notation $\big|_{\ldots}$ applies to the whole expression. 
These differ from the standard slow-roll approximation in that $\delta_{p+1}$ terms 
are not dropped just because a lower-order $\delta_{p}$ appears.

Likewise the tensor observables become
\begin{eqnarray}
\ln \Delta_{+,\times}^2&\approx &   \ln \frac{H^2}{2\pi^2} - \frac{8}{3} \epsilon_H \Big|_{x= x_1}
, \nonumber\\
n_t &  \approx& -2 \epsilon_H \Big|_{ x= x_1}, \nonumber\\
\alpha_t &\approx&  -4\epsilon_H^2 - 4 \epsilon_H\delta_1 \Big|_{ x= x_1} , \qquad ({\rm OLO}).
\label{eqn:OSRt}
\end{eqnarray}
These expressions for $n_t$ and $\alpha_t$ are the same as the standard slow-roll expressions due to the lack of a running of $n_t$ at linear order in
$\epsilon_H$ and $\delta_n$.

Since OLO self-consistently accounts for both running of the tilt
and running of the running of the tilt by absorbing the next correction into the evaluation point, \ it is both a simpler and better approximation than keeping all  $O(1/N^2)$ and $O(1/N\Delta N)$ terms in the standard slow-roll approximation 
as often used in the literature
(e.g.~\cite{Planck:2013jfk,Ade:2015lrj,Baumann:2014cja}; see  \S \ref{sec:monodromy} for tests).  Note also that the standard approach is often phrased in terms of the 
Hubble flow parameters where $\delta_p$ and its hierarchy equation,
Eq.~(\ref{eqn:hubbleeom}),  is replaced by 
\begin{equation}
\frac{d\ln \epsilon_{p}}{dN} = \epsilon_{p+1} ,
\label{eqn:Hubbleflow}
\end{equation}
so that 
\begin{eqnarray}
\epsilon_1 &\equiv& \epsilon_H ,\nonumber\\
\epsilon_2 &=& 2(\epsilon_H  + \delta_1), \nonumber\\
\epsilon_2\epsilon_3 &=& 
4\epsilon_H^2  + 6\epsilon_H\delta_1 - 2\delta_1^2 + 2\delta_2 .
\label{eqn:Hubbleepsp}
\end{eqnarray}
While the two series are algebraically equivalent when all terms are kept, Hubble flow parameters are inconvenient
for cases that lack a strong hierarchy [see  Eq.~\eqref{eqn:HSO} and Fig.~\ref{fig:DN3nsalphah}].   
Equation~(\ref{eqn:Hubbleflow}) says that for $\epsilon_{p+1}$ to remain small, the fractional change in $\epsilon_p$ over an efold must be small. 
Hence order counting in powers of $\epsilon_p$
automatically conflates $\Delta N$ and $N$ as Eq.~(\ref{eqn:Hubbleepsp}) illustrates.   
For example in the cases considered in \S \ref{sec:monodromy} where the variations occur on $\Delta N\ll 1$, these parameters would have
poles, whereas the $\delta_p$ do not.

\subsection{Potential Slow Roll Parameters}
\label{sec:potential}

We can likewise relate the $G^{(p)}$ hierarchy to the equivalent in derivatives of the potential.
Using the exact  background equations of motion for the inflaton and the expansion
\begin{eqnarray}
(3+\delta_1)H^2  \frac{d \phi}{dN} &=& -V^{(1)}, \nonumber\\
(3-\epsilon_H) H^2 &=& V ,
\label{eqn:background}
\end{eqnarray}
we can relate the derivatives of the potential to the Hubble slow-roll parameters
\begin{eqnarray}
\U &\equiv & \mk{\f{V^{(1)}}{V}}^2 = 2\epsilon_H \frac{ (1+\delta_1/3)^2}{(1-\epsilon_H/3)^2}, \nonumber\\
\sV_1 &\equiv & \frac{V^{(2)}}{V} = \frac{\epsilon_H -\delta_1 -\delta_2/3}{1-\epsilon_H/3}.
\label{eqn:U}
\end{eqnarray}
Note that to lowest order Eq.~(\ref{eqn:U}) gives $\U \approx 2\epsilon_H$ and
$\sV_1 \approx \epsilon_H-\delta_1-\delta_2/3$.   As shown in the Appendix, these
imply $G' \approx  3\U - 2\sV_1$.  Likewise
the series  of higher-order derivatives with respect to efolds is replaced by
higher-order derivatives of the potential with respect to the field
\be \sV_p \equiv \left( \frac{V^{(1)}}{V} \right)^{p-1}  
\frac{ V^{(p+1)}}{V},
\ee
where $V^{(p)} \equiv d^p V/d \phi^p$.    
In the literature the first few parameters
are also known as $\epsilon_V = \U/2$, $\eta_V=\sV_1$, and $\xi_V=\sV_2$.
Terms that are quadratic in $\U, \sV_p$ 
are higher order in the deviation from a de Sitter expansion.
However, the relative size of the $\sV_p$ parameters  depends on the smoothness of 
the potential $V(\phi)$ just as that of $\delta_p$ depends on the smoothness of 
$H(N)$.

We explicitly carry out these conversions in the Appendix  and outline the results
relevant to the ${\cal O}(1/N)$ expansion here.
Note that to leading order 
Eq.~(\ref{eqn:background}) provides the condition for
a friction dominated roll or attractor solution,
\begin{equation}
\frac {d\phi}{dN } \approx -\frac{1}{3} \frac{V^{(1)}}{H^2} \approx -\frac{V^{(1)}}{V},\label{eqn:dpdn}
\end{equation}
which also converts the two senses of derivatives
\begin{equation}
\frac{d}{d\ln\eta}  \approx \frac{V^{(1)}}{V} \frac{d}{d\phi},
\end{equation}
or
\begin{equation}
\frac{d \U }{d\ln\eta} \approx 0, \quad
\frac{d\sV_{p}}{d\ln\eta} \approx \sV_{p+1},
\end{equation}
and thus 
\begin{equation}
G^{(p)} \approx - 2\sV_p + {\cal O}(N^{-2}),
\end{equation}
for $p\geq 2$.   The Taylor series in $G^{(p)}$ has a one-to-one relationship to the same
series in $\sV_p$ (see the Appendix for details).
The convergence criteria for the Taylor expansion in Eq.~(\ref{eqn:convergence}) likewise
becomes a direct condition on the potential slow-roll parameters
\begin{equation}
\lim_{p\rightarrow \infty} \left|  \frac{G^{(p)}}{G^{(p-1)}} \right|   = \lim_{p\rightarrow \infty} \left|  \frac{\sV_p}{\sV_{p-1}}  \right| < 2.
\label{eqn:convergenceV}
\end{equation}
We can interpret this condition by approximating
\begin{equation}
\frac{\sV_p}{\sV_{p-1}} \approx - \frac{d\phi}{dN } \frac{V^{(p+1)}}{V^{(p)} } \approx -\frac{d \ln V^{(p)}}{d N }.
\end{equation}
Thus the convergence condition is that the fractional variation in $V^{(p)}$ experienced by 
the field across an efold
is small.

Although the OLO approximation again follows directly from evaluating $G^{(p)}$ in terms 
of these parameters, it is useful to restore the general evaluation epoch $\ln x_f$ to 
clarify its relationship to the field position on the potential.
Equations~(\ref{eqn:Gseries}) and (\ref{eqn:tiltrun}) become 
\begin{align}
\Delta^2_\curv &\approx \f{V}{12\pi^2 {\cal U}}\left[ 1+\mk{3\tq_1-\f{7}{6}} {\cal U} 
- 2\sum_{p=1}^{\infty}\tq_p \sV_p  \right] , \nonumber\\
n_s-1 &\approx -3{\cal U}  + 2 \sV_1 + 2\sum_{p=1}^{\infty}\tq_p \sV_{p+1} , \nonumber \\
\alpha& \approx -2 \sV_2 
-2\sum_{p=1}^{\infty}\tq_p \sV_{p+2} -6 \U^2 + 8 \U \sV_1  ,
\label{eqn:tiltrunV}
\end{align}
where the potential terms are evaluated at the field position $\phi(\ln x_f)$ which we
will write as $\ln x_f$ as shorthand notation.
In this language, optimization of $\ln x_f$ is equivalent to optimization of the
field position at which to evaluate the potential parameters.  
From the field position $\ln x_f=\ln x_0=0$, a shift in evaluation point to
$\ln x_1=\tq_1(0)$ causes a shift of
\begin{equation}
\delta \phi \approx -  \frac{ V^{(1)} }{V}\delta  N  \approx   \frac{ V^{(1)}}{V} \tq_1(0)
\end{equation}
or a change in potential parameters
\begin{align}
\delta V &\approx  V^{(1)} \delta \phi \approx \U V \tq_1(0), \nonumber \\
\delta \U &\approx 2 \frac{V^{(1)}}{V} ( \sV_1 - \U )  \delta \phi
 \approx 2 \U (\sV_1-\U) \tq_1(0), \nonumber \\
\delta \sV_p &\approx\frac{V}{V^{(1)}} \sV_{p+1}\delta \phi \approx \sV_{p+1} \tq_1(0)
\end{align}
that cancel the NO corrections in the hierarchy.  
In fact starting from an arbitrary $\ln x_f$ we can define 
\begin{align}
\tilde \sV_p(\ln x_f) &\equiv \sV_p(\ln x_f) + \tq_1(\ln x_f)  \sV_{p+1}(\ln x_f) \nonumber\\
&\approx  \sV_p(\ln x_1) .
\label{eqn:Vtilde}
\end{align}
For the OLO 
choice, $\tilde \sV_p(\ln x_1)=\sV_p(\ln x_1)$, and for alternate $\ln x_f$, $\tilde \sV_p(\ln x_f)$ differs
from $\sV_p(\ln x_1)$ only by next-to-NO corrections.   $\tilde \sV_n$ will be useful 
for comparing higher-order approximations in \S \ref{sec:second}.
In fact for an arbitrary order
we could define $\tilde \sV_p \equiv \sV_p + \sum \tq_n \sV_{n+p}$ to absorb the appropriate number of correction
terms into an effective potential parameter.

Thus the OLO approximation simply amounts to shifting the evaluation point in field space by the same amount for all observables.  Maintaining this consistency
condition is important in testing models.  
Aside from this evaluation point, the OLO approximation ($\ln x_f= \ln x_1$)
takes the same form as the standard LO approximation ($\ln x_f=0$),
\begin{align}
\Delta^2_\curv &\approx \f{V}{12\pi^2 {\cal U}}\left( 1 -\f{7}{6} {\cal U} \right),\nonumber\\
n_s-1 &\approx -3{\cal U}  + 2 \sV_1  , \nonumber \\
\alpha &\approx -2 \sV_2   -6 \U^2 + 8 \U \sV_1  , \qquad {\rm (O/LO)}.
\label{eqn:tiltrunVopt}
\end{align}
Though identical in form, the standard approximation  assumes $\sV_2$ is  ${\cal O}(N^{-2})$
making $\alpha$ suppressed with respect to $n_s-1$ and absent in its leading-order prediction, 
not because of resummation to a new evaluation point but because of its truncation at ${\cal O}(N^{-1})$.
Again, the advantage of this OLO  is that not only is it as simple as the standard LO slow-roll
prescription, but it consistently incorporates the NO corrections
that run the tilt and run the running of the tilt. 
Corrections to it
are suppressed by  ${\cal O}(1/\Delta N^2)$.

We shall see that the same is not true
for approaches that correct for evolution at a fixed order in powers of $\Delta N\sim N$ 
as is often used in the literature
(e.g.~\cite{Lidsey:1995np,Gao:2014yra,Boyle:2014kba}).  For instance
if $\alpha$ is of order $n_s-1$, the standard second-order approach is really only appropriate for
strictly constant $\alpha$ or equivalently a purely cubic potential where the cubic term
is large but no higher terms are important. 
Importantly, in OLO 
we absorb the effect of the quartic term $\sV_3$ in the optimized evaluation of $\alpha$ to correct for the
running of its value between horizon crossing and freeze-out.   
In the second-order approach,
tilt is run to the freeze-out point, but running is not.    
One would then infer an incorrect relationship between the derivatives
of the potential and hence potentially incorrectly falsify a true model that is not purely cubic from observations.
Thus, a second-order approach is both more complicated
and less general than the OLO approach. 
We illustrate this in \S \ref{sec:lowfrequency} for the monodromy example.

For tensor fluctuations, conversion of $G^{(p)}_h$  to the potential parameters gives
\begin{eqnarray}
\Delta_{+,\times}^2(k) &\approx&  \frac{V}{6\pi^2} \left[ 1- \frac{7}{6}\U \right],
\nonumber\\
r &\approx& 8 \U , \nonumber\\
n_t &\approx& -\U  , \nonumber\\
\alpha_t &\approx& -2 \U^2 + 2\U \sV_1,  \qquad {\rm (O/LO)}
\label{eqn:OLOtensor}
\end{eqnarray}
where the only NO correction is in the power spectrum itself,
\begin{eqnarray}
\delta \Delta_{+,\times}^2(k) =  \frac{V}{6\pi^2} \tq_1 \U , \qquad ({\rm NO}).
\end{eqnarray}
Here again OLO has the same accuracy as NO since $\tq_1(\ln x_1)=0$.
In particular the consistency relation $r\approx -8n_t$ remains unchanged 
regardless of the evaluation point $\ln x_f$.

Finally, the OLO approximation provides a simple inverse relationship between the
observables $r$, $n_s$ and $\alpha$ and the potential parameters
\begin{eqnarray}
\U &\approx& \frac{r}{8}, \nonumber\\
\sV_1 &\approx & \frac{n_s-1}{2}+ \frac{3r}{16} ,  \nonumber\\
\sV_2 &\approx&  -\frac{\alpha}{2} + \frac{1}{4}(n_s-1) r 
 + \frac{3r^2}{64}, \qquad  {\rm (OLO)}.
 \label{eqn:reconOLO}
\end{eqnarray}
Even if $r$ is not accurately measured, strong upper limits
where $r \ll 8|1-n_s|/3$ allow a reconstruction of
$\sV_1$ and $\sV_2$.

Equivalently, those observables allow a local reconstruction of the potential to cubic order around the 
field value where $x(\phi_0)=x_1$
\begin{equation}
\frac{V}{V_0} \approx 1 + \sqrt{\U} (\phi-\phi_0) + \frac{\sV_1}{2}  (\phi-\phi_0)^2 + \frac{\sV_2}{6\sqrt{\U}} (\phi-\phi_0)^3.
\label{eqn:cubic}
\end{equation}
The potential amplitude $V_0$, or the energy scale of inflation, can likewise be determined from measuring $\Delta_{+,\times}^2$ or equivalently $\Delta_\curv^2$ and $r$ as is well known.

\section{Optimized Next- and Second-Order Approximation}
\label{sec:second}

The OLO approximation of the previous section is accurate
at first order in the deviations from scale invariance ${\cal O}(1/N)$
and next-to-leading order (NO) in a hierarchy of slow-roll parameters separated by $\Delta N$, the temporal scale of features during inflation.   
For cases where $\Delta N \sim 1$ and the hierarchy is weakly convergent
or if higher accuracy is desired for future observations, we can generalize this approach.
In \S \ref{sec:SOGSR} we consider next-to-NO in the hierarchy 
and an ${\cal O}(1/N^2)$  in the de Sitter deviations.   
In \S \ref{sec:ONO}, we develop 
a new optimized NO (ONO) approximation 
that retains the simplicity of the NO approximation. 
This also allows
us to contrast OLO and ONO with the standard second order (SO) approach which conflates
$\Delta N$ and $N$ in \S \ref{sec:SO}.

\subsection{General Expression}
\label{sec:SOGSR}

In terms of the hierarchy of slow-roll parameters,
the next-to-NO approximation involves keeping terms to order $p=2$ in the
Taylor expansion of $G$ in  Eq.~(\ref{eqn:Gseries})  
\begin{equation}
I_0 \approx G(\ln x_f)  + \sum_{p=1}^{2} \tq_p  G^{(p)}(\ln x_f).
\label{eqn:Gseriesxf}
\end{equation}
To also include all ${\cal O}(1/N^2)$ effects, we  evaluate the 
second-order terms $I_1$ and $I_2$ from 
Eq.~(\ref{eqn:I1I2}).   In these corrections, $G'$ and $f'/f$ can be taken as constants
evaluated at $\ln x_f$ as their evolution introduces terms of
${\cal O}(1/N^2\Delta N)$ which we omit.  Combining these pieces, we obtain
\begin{eqnarray}
\ln \Delta_\curv^2(k) &\approx& G(\ln x_f)  +  \tq_1 G'(\ln x_f)  + \tq_2  G''(\ln x_f) \nonumber\\
&& + \frac{\pi^2}{8} [G'(\ln x_f)]^2 -4 \left[\frac{f'}{f}(\ln x_f)\right]^2  .
\label{eqn:Gseries2}
\end{eqnarray}
The same expression is valid for tensors with the appropriate replacements as before.

Since the variation in $\ln k$ of the last two $I_1$ and $I_2$ terms in Eq.~(\ref{eqn:Gseries2}) is
itself ${\cal O}(1/N^2\Delta N)$, we do not include their impact on $n_s$ or $\alpha$.   The only
${\cal O}(1/N^2)$ terms that appear for those quantities come from the nonlinear
relationship between $G^{(p)}$ and slow-roll parameters.   This justifies our
inclusion of such terms in $\alpha$ in the OLO approximation since there are no
further corrections from intrinsically second-order GSR effects.

As shown in the Appendix, 
in terms of the potential slow-roll parameters we obtain
\begin{eqnarray}
\Delta^2_\curv & \approx & \frac{V}{12\pi^2\U} \Big[ 1 +  
\left( 3\tq_1 -\frac{7}{6} \right) \U  - 2 \tq_1 \sV_1 - 2\tq_2 \sV_2 
\nonumber\\ 
&& + \left( 3 \tq_2 - \frac{2}{3} \tq_1 - \frac{103}{9} + \frac{3\pi^2}{2} \right) \U^2 
\nonumber\\
&& + \left( -4 \tq_2 + \frac{2}{3} \tq_1 + 15 - 2 \pi^2 \right) \U \sV_1
\nonumber\\
&& + \left( 4 \tq_2 - \frac{2}{3} \tq_1 - \frac{13}{3} + \frac{2\pi^2}{3} \right) \sV_1^2 \Big] ,
\nonumber\\
n_s & \approx & 1 - 3\U + 2\sV_1 + 2 q_1\sV_2 + 2q_2 \sV_3 
\nonumber\\ 
&& + \left( 6 \tq_1 - \frac{17}{6} \right) \U^2  -\left( 8 \tq_1 - \frac{5}{3} \right) \U \sV_1+ \frac{2}{3}  \sV_1^2  , 
\nonumber\\
\alpha & \approx & -2 \sV_2 - 2\tq_1 \sV_3 - 2\tq_2 \sV_4 - 6\U^2 + 8 \U \sV_1 ,
\label{eqn:NO}
\end{eqnarray}
for the  scalar expression and
\begin{eqnarray}
\Delta^2_{+,\times} & \approx & \frac{V}{ 6\pi^2} \Bigg[ 1+
\left( \tq_1 - \frac{7}{6} \right) \U 
\nonumber\\
&& +\left( - \tq_2 + \frac{4}{3} \tq_1 - \frac{8}{3} + \frac{\pi^2}{6} \right) \U^2
\nonumber\\
&& + \left( 2 \tq_2 - 2 \tq_1 + \frac{17}{9} \right) \U \sV_1 \Bigg] , \nonumber\\
r & \approx & 8\U -16 \tq_1 \U(\U -\sV_1) ,  \nonumber\\
n_t & \approx & -\U + \left( 2 \tq_1 - \frac{5}{2} \right) \U^2 - 2 ( \tq_1 -1 ) \U \sV_1 ,
\nonumber\\
\alpha_t & \approx & -2\U(\U - \sV_1),
\label{eqn:NOt}
\end{eqnarray}
for the tensor expression.  
These are the master equations for all next-to-NO  in the hierarchy and 
second order in $1/N$ approximations.  
They of course also include lower-order
approximations such as LO, OLO and SO with the appropriate zeroing of terms.

\subsection{Optimized Evaluation}
\label{sec:ONO}

We can again optimize the evaluation epoch $\ln x_f$ to
make the optimized NO approximation as
accurate as a general next-to-NO approximation.
We therefore take $\tq_2(\ln x_2)=0$ in
Eq.~(\ref{eqn:q2q3})
and pick the solution that  minimizes the next correction $|\tq_3|$,
\begin{eqnarray} 
\tq_2(\ln x_2) &\equiv & 0, \nonumber\\
\tq_1(\ln x_2) &=& \frac{\sqrt{3\pi^2 -4} }{6}\approx 0.84 ,\nonumber\\
\ln x_2 &=&  \ln x_1 - q_1(\ln x_2)  \approx 0.22,  \quad {\rm (ONO)}.
\end{eqnarray}
We call the evaluation of
Eqs.~(\ref{eqn:NO}) and (\ref{eqn:NOt}) with these values the optimized next-order (ONO) approximation.    
Note that this has the effect of zeroing the highest 
$\sV_p$ term in each observable.

For a strong hierarchy $\Delta N\gg 1$, despite the  shift in the actual evaluation point
to $\ln x_2$,
observables remain effectively evaluated at $\ln x_1$ since 
\begin{eqnarray}
I_0 & = & G(\ln x_2) + \tq_1(\ln x_2) G'(\ln x_2) + \ldots \nonumber\\
&= & G(\ln x_2) + ( \ln x_1 -\ln x_2 ) G'(\ln x_2)+\ldots \nonumber\\
&= & G(\ln x_1) - \frac{1}{2}  ( \ln x_1 -\ln x_2 )^2 G''(\ln x_1) + \ldots  \nonumber\\
&= & G(\ln x_1) + q_2(\ln x_1) G''(\ln x_1)+ \ldots
\end{eqnarray}
where $\ldots$ represents $G^{(3)}$ terms and higher.
Evaluation at $\ln x_2$ simply increases the accuracy by including the first
correction to this approximation.  Since $\ln x_1$ is a maximum of $|q_2|$ this correction can become important for $\Delta N \sim 1$.  Furthermore $\ln x_2$ is a minimum of
$|q_3|$ and so even the $G^{(3)}$ error is minimized in ONO (see Table~\ref{tab:optimized}).

In terms of  the reconstruction of $\U, \sV_1, \sV_2$ from $n_s,\alpha,r$ the system
is no longer closed due to the appearance of $q_1 \sV_3$ in $\alpha$.   As in the case of
$I_0$ above, this term represents the leading-order effect of a shift in the evaluation of $\sV_2$ back to $\phi(\ln x_1)$  from $\phi(\ln x_2)$.  We can therefore define instead the
reconstructed parameters as the $\tilde \sV_p$ of Eq.~(\ref{eqn:Vtilde})
which can equally well be predicted from  any given model for comparison.  Moreover
in the quadratic terms the difference between employing $\sV_p$ and $\tilde \sV_p$ is
at most ${\cal O}(1/N^2\Delta N)$ which we neglect.  Thus the master equation can be written
entirely in terms of $\tilde \sV_p$.      Iterating the inversion we obtain
\begin{eqnarray}
\U &\approx& \frac{r}{8} -\frac{q_1}{8}  (n_s-1)r - \frac{\tq_1}{64} r^2   , \nonumber\\
\tilde\sV_1 &\approx & \frac{n_s-1}{2}+ \frac{3r}{16}-\frac{(n_s-1)^2}{12}
\nonumber\\ &&
+\left(\frac{\tq_1}{16}-\frac{11}{96} \right)(n_s-1)r
+\left(\frac{3\tq_1}{128}-\frac{7}{768} \right)r^2 ,  \nonumber\\
\tilde \sV_2 &\approx&  -\frac{\alpha}{2} + \frac{1}{4}(n_s-1) r 
+ \frac{3r^2}{64}, \qquad  {\rm (ONO)},
\label{eqn:reconONO}
\end{eqnarray}
for the potential reconstruction.

\subsection{Standard Second Order}
\label{sec:SO}

We can also contrast the OLO and ONO approximations with the standard second-order approach.  The standard second-order approach is not simply a choice of $\ln x_f=0$ in the master equations.  By  conflating  $\Delta N$ with $N$, it truncates at a fixed order in both.   In other words it assumes $\sV_p = {\cal O}(1/N^p)$
and so in the  ${\cal O}(1/N^2)$ expansion,
\begin{eqnarray}
\sV_{3},\sV_4,\ldots &=& 0, \qquad {\rm (SO)}, \nonumber\\
\tq_1 \approx 1.06, \, \tq_2 &\approx& 0.21 
\end{eqnarray} 
in Eqs.~(\ref{eqn:NO}) and (\ref{eqn:NOt}).  Thus, the standard second-order
(SO) approximation implicitly assumes that the potential is cubic in form (see the Appendix
for the relationship to the second-order Hubble flow approximation).
Since truncation at 
a fixed $\sV_p$ is not the same as truncation at a fixed order in the Taylor expansion,
this means that different observables are effectively evaluated at an inconsistent
$\ln x_f$ or field position.   
We shall see that this makes SO an even worse approximation than OLO for cases where
$\Delta N \ll |N|$ and the potential is not cubic.

In particular, dropping $q_1 \sV_3$ terms in $\alpha$ but keeping $q_1 \sV_2$ terms
in $n_s-1$ leads to the reconstruction
\begin{eqnarray}
\U &\approx& \frac{r}{8} -\frac{q_1}{8}  (n_s-1)r - \frac{\tq_1}{64} r^2   , \nonumber\\
\sV_1 &\approx & \frac{n_s-1}{2}+ \frac{3r}{16}+ \frac{\tq_1}{2}\alpha -\frac{(n_s-1)^2}{12}
\nonumber\\ &&
-\left(\frac{3\tq_1}{16}+\frac{11}{96} \right)(n_s-1)r
-\left(\frac{3\tq_1}{128}+\frac{7}{768} \right)r^2 ,  \nonumber\\
\sV_2 &\approx&  -\frac{\alpha}{2} + \frac{1}{4}(n_s-1) r 
+ \frac{3r^2}{64}, \qquad  {\rm (SO)},
\label{eqn:reconSO}
\end{eqnarray}
which is not simply a redefinition of $\tq_1$ in Eq.~(\ref{eqn:reconONO}).

In particular $\alpha$ now appears in the formula for $\sV_1$.
For a purely cubic potential, these differences simply reflect a different expansion point $\phi_0$ for the 
reconstruction in Eq.~(\ref{eqn:cubic}).   However for a potential that contains quartic corrections, the
SO reconstruction can lead to inconsistencies between the parameters $\{ \U,\sV_1,\sV_2 \}$ that do not simply
reflect evaluation at a different field value for the potential.
We now turn to an illustrative example of such a model.

\section{Monodromy Case Study}
\label{sec:monodromy}

Axion monodromy inflation  \cite{Silverstein:2008sg} provides a well-motivated theoretical
model where the various approximation schemes discussed in the previous sections can be tested.  
The monodromy potential is  the sum of a smooth and oscillatory part
\begin{eqnarray}
V(\phi) &=& \bar V(\phi) + \delta V(\phi),
\end{eqnarray}
for which we take the simplest case
\begin{eqnarray}
\bar V(\phi)&=& \lambda \phi ,\nonumber\\
\delta V(\phi)&=& \Lambda^4 \cos \mk{\frac{\phi}{f}+\theta} .
\end{eqnarray}
Depending on the frequency of this oscillation, the curvature power
spectrum may gain a  large local running of the tilt \cite{Kobayashi:2010pz,Meerburg:2014bpa} or high
frequency features for which the slow-roll parameters have an inverted hierarchy and lose their ability to characterize observables directly \cite{Flauger:2009ab}.

In \S \ref{sec:lowfrequency}, we study the low frequency case and test the accuracy of the LO,
OLO, ONO and SO approximations.   We develop optimized GSR approximations for
the high frequency case in \S \ref{sec:highfrequency}, extending results of Ref.~\cite{Flauger:2009ab} for second-order corrections in the large amplitude limit as well as establishing the
connection of the running amplitude, frequency, and phase of the oscillation to parameters of the potential.

\begin{figure}[t]
\centering
\includegraphics[width=3.5in]{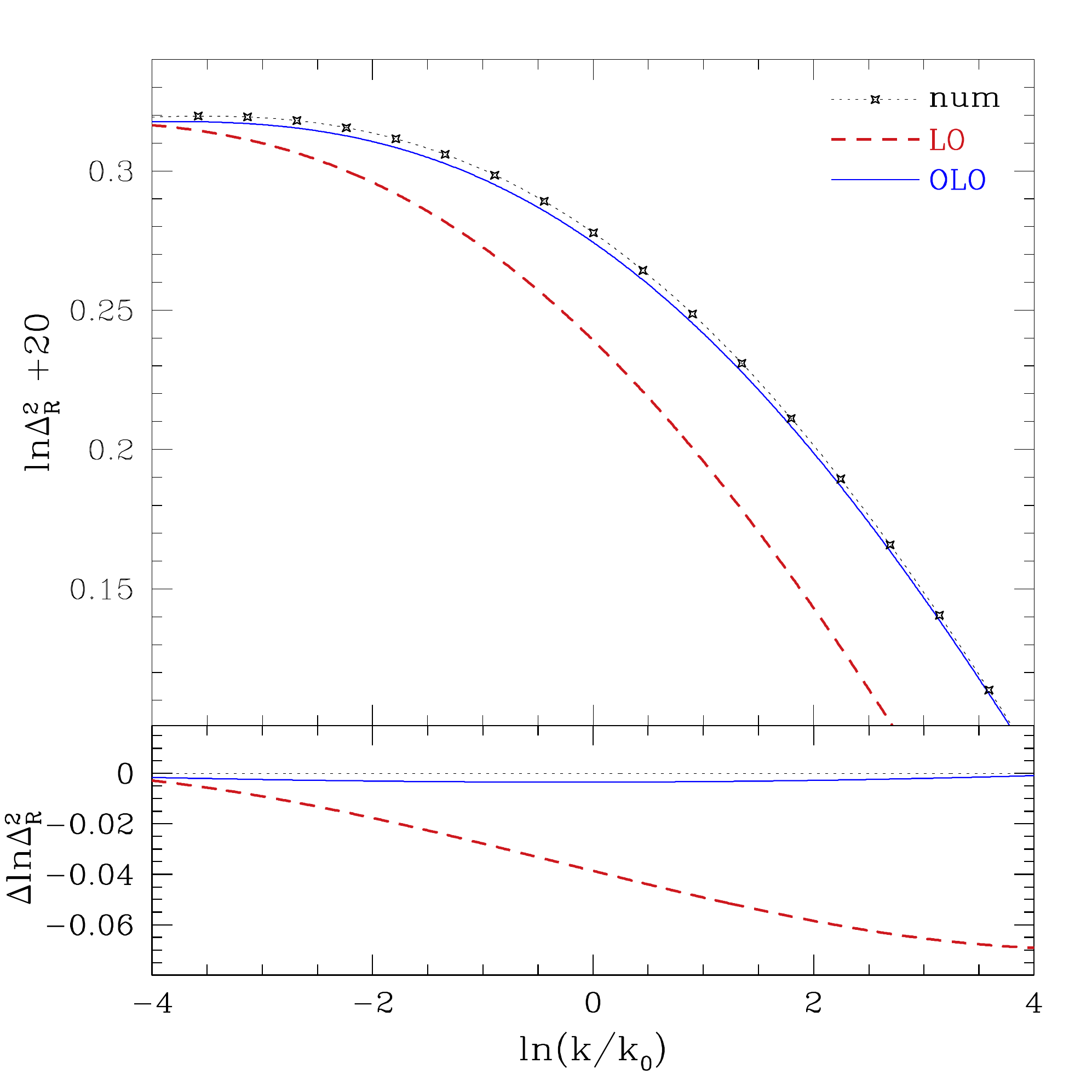}
\caption{Low frequency $\omega=1/3$ oscillations
  in the curvature power spectrum
  $\Delta_\curv^2$ under the leading-order (LO) and optimized leading-order (OLO) 
  approximations.  While the LO approximation introduces a large error in the power spectrum
  compared to numerics, the OLO results show that they can be corrected by 
  simply shifting the relationship between $\ln k$ and $\phi$.   Here the amplitude of the oscillation is set to $\alpha_{\rm max}=-0.01$.}
\label{fig:DN3pow}
\end{figure}

\begin{figure}[t]
\centering
\includegraphics[width=3.5in]{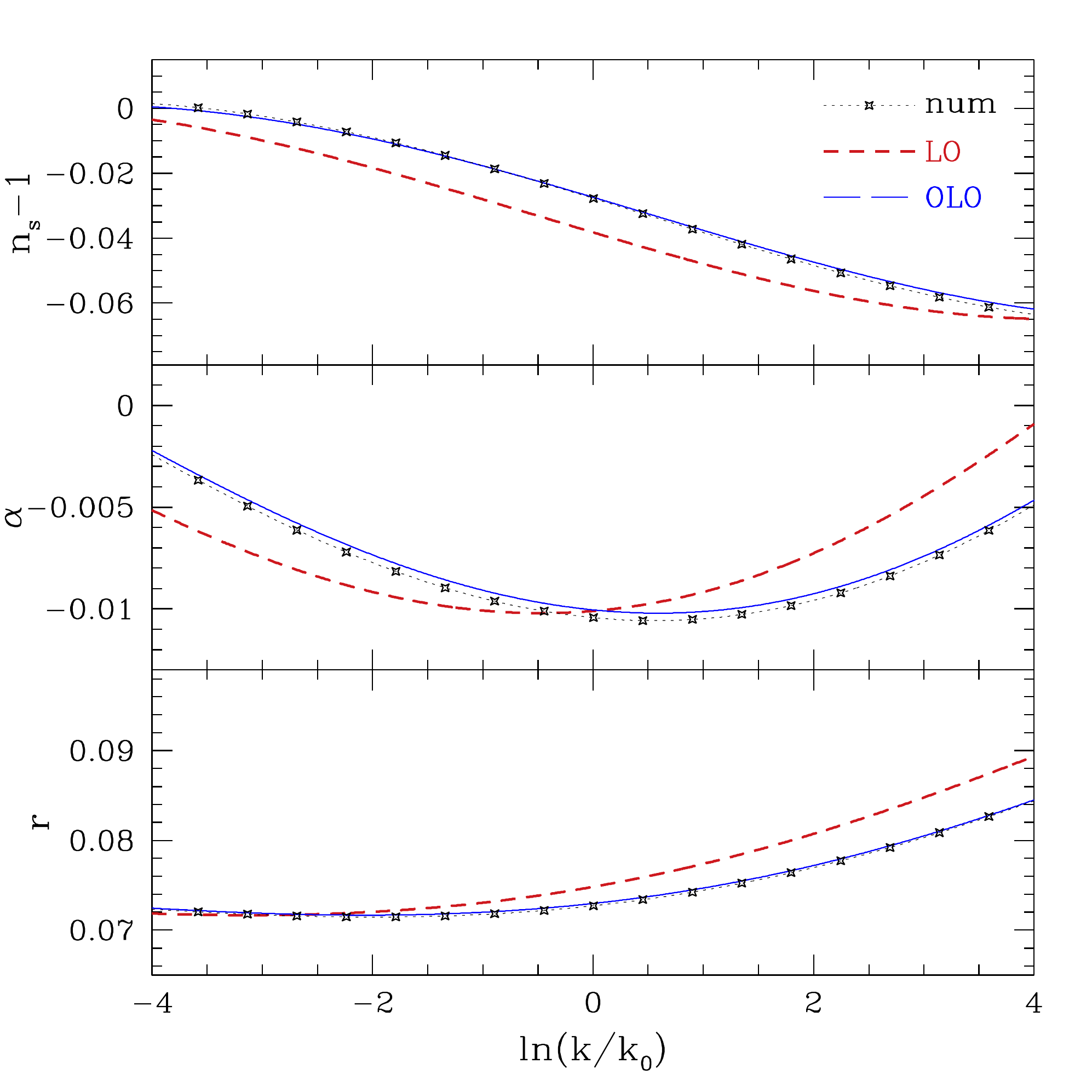}
\caption{Low frequency $\omega=1/3$ oscillations
  in $n_s$, $\alpha$ and $r$  under the LO and OLO
  approximations as in Fig.~\ref{fig:DN3pow}.   The OLO approximation again provides an
  excellent approximation that corrects the shift in the LO results
  even in this case where $\omega$, the hierarchical separation between
  $\alpha$ and $n_s-1$, is not much less than unity. }
\label{fig:DN3obs}
\end{figure}

\begin{figure}[t]
\centering
\includegraphics[width=3in]{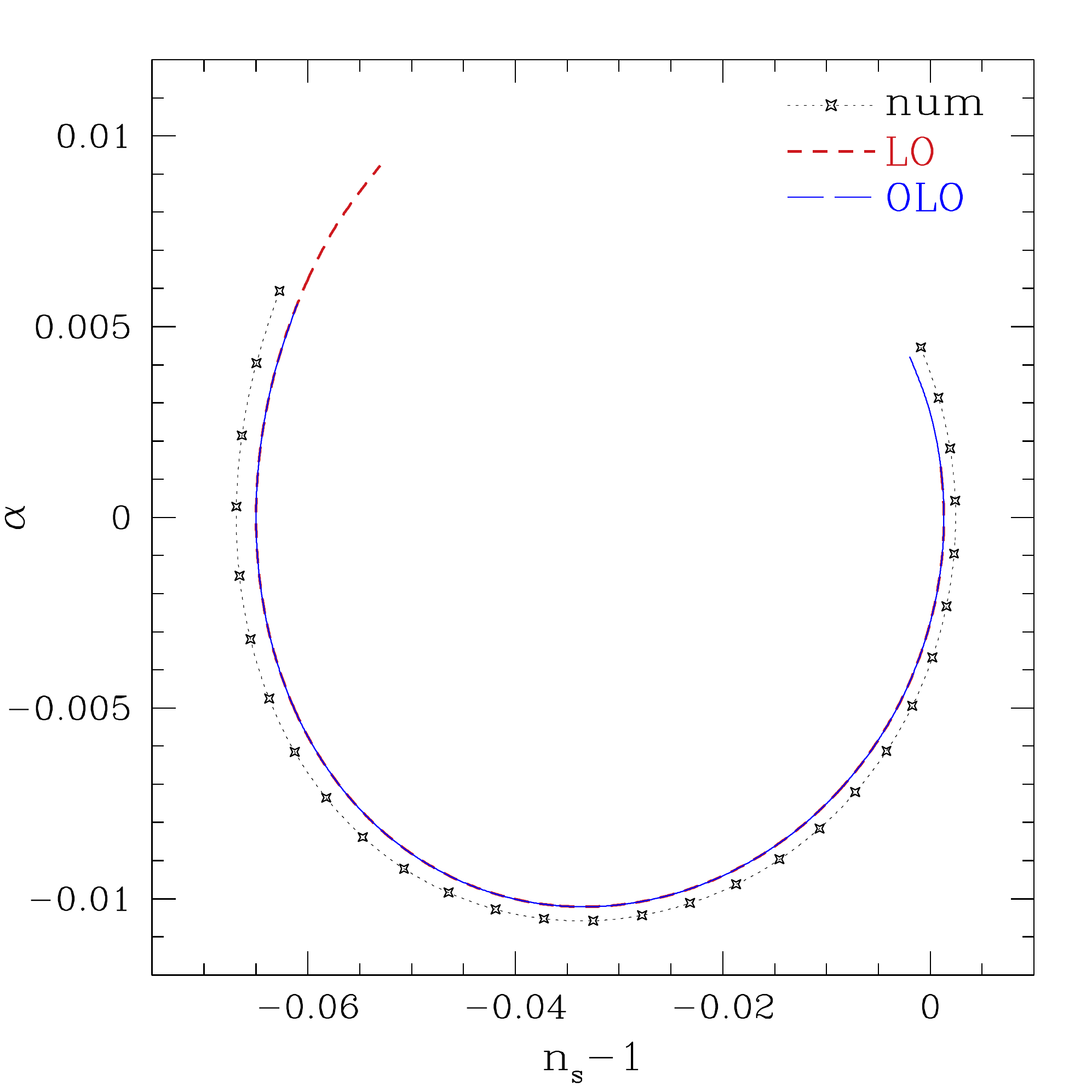}
\caption{Low frequency $\omega=1/3$ oscillation trajectories in the
  $n_s$-$\alpha$ plane under the LO and OLO
  approximations as in Fig.~\ref{fig:DN3pow}.   Both LO and OLO trajectories
  are good approximations to the numerical results emphasizing that the LO results
  differ simply in the evaluation point along the trajectory, parametrically evaluated here
  for the same range in observed wavenumber $k$. }
\label{fig:DN3nsalpha}
\end{figure}

\subsection{Low Frequency Oscillations}
\label{sec:lowfrequency}

In the low frequency  regime, the slow-roll parameters are hierarchical
and yet the running of tilt can be large and nearly constant across the well-measured
few efolds of the cosmic microwave background (CMB) and large-scale structure.
In this case, the OLO and ONO
slow-roll expressions  should provide a good approximation for observables whereas we have
argued that the standard SO expressions do not.

For tests of these approximations, we choose the parameters of the potential such that the smooth piece provides a power
spectrum amplitude and tilt of the right order at some fiducial scale $k_0$, 
\begin{eqnarray}
\bar \Delta_\curv^2(k_0) &=& \bar A_s = 2.5 \times 10^{-9} ,\nonumber\\
\bar n_s  &=& 0.97,
\end{eqnarray}
which determines
\begin{eqnarray}
\phi_0 &=& \sqrt{\frac{3}{1-\bar n_s} }  ,\nonumber\\
\lambda &=& \frac{12 \pi^2}{\phi_0^3}\bar A_s.
\end{eqnarray}
As shown by the OLO approximation, fluctuations freeze out at
\begin{equation}
{k_0}\eta(\phi_0) =x_1  \approx 2.89,
\end{equation}
which provides a conversion between $k_0$ and $\phi_0$ which we use as a definition throughout.
The parameter $f$ determines the frequency of the oscillation and thus the separation 
between terms in the slow-roll hierarchy.  Since
\begin{equation}
\left| \frac{{\cal V}_p}{{\cal V}_{p-1}} \right|  \approx \frac{1}{f\phi_0} =\frac{{| d\phi/d N}|}{f} \equiv \omega \ll 1 ,
\end{equation} 
we fix $f$ by a choice of $\omega \sim \Delta N^{-1}$.    
Given Eq.~(\ref{eqn:convergenceV}), the series expansion in $p$ converges 
if $\omega < 2$. 
Thus in the low frequency regime $\omega \ll 1$, the optimized slow-roll approach
should be a good approximation.

We set the phase of the oscillation so that $\alpha$ is at an extrema $\alpha_{\rm max}$ at $k_0$,
\begin{equation}
\theta= \frac{\pi}{2} - {\rm Mod}(\phi_0/f,2\pi),
\end{equation}
for low frequency oscillations.
Finally we set the amplitude of the oscillations $\Lambda^4$ by relating
it to the running of the tilt at the $k_0$ extrema 
\begin{equation}
\Lambda^4 = - \frac{1}{2}\lambda \phi_0^2 f^3 \alpha_{\rm max}.
\end{equation}
In practice
we choose $\alpha_{\rm max} = -0.01$ in our examples.  Since $\Lambda^4$ increases as $\omega \rightarrow 0$ at fixed $\alpha_{\rm max}$
and $n_s-1$,  higher-order terms in $\delta V$ eventually become important.  
Therefore, we use $\alpha_{\rm max}$ simply as a proxy for choosing $\Lambda^4$ 
rather than associating it with the actual value of $\alpha$ at $k_0$.   
Likewise, $\bar A_s$ and $\bar n_s$ also receive higher-order corrections 
and are used here simply to set parameters.

\begin{figure}[t]
\centering
\includegraphics[width=3.5in]{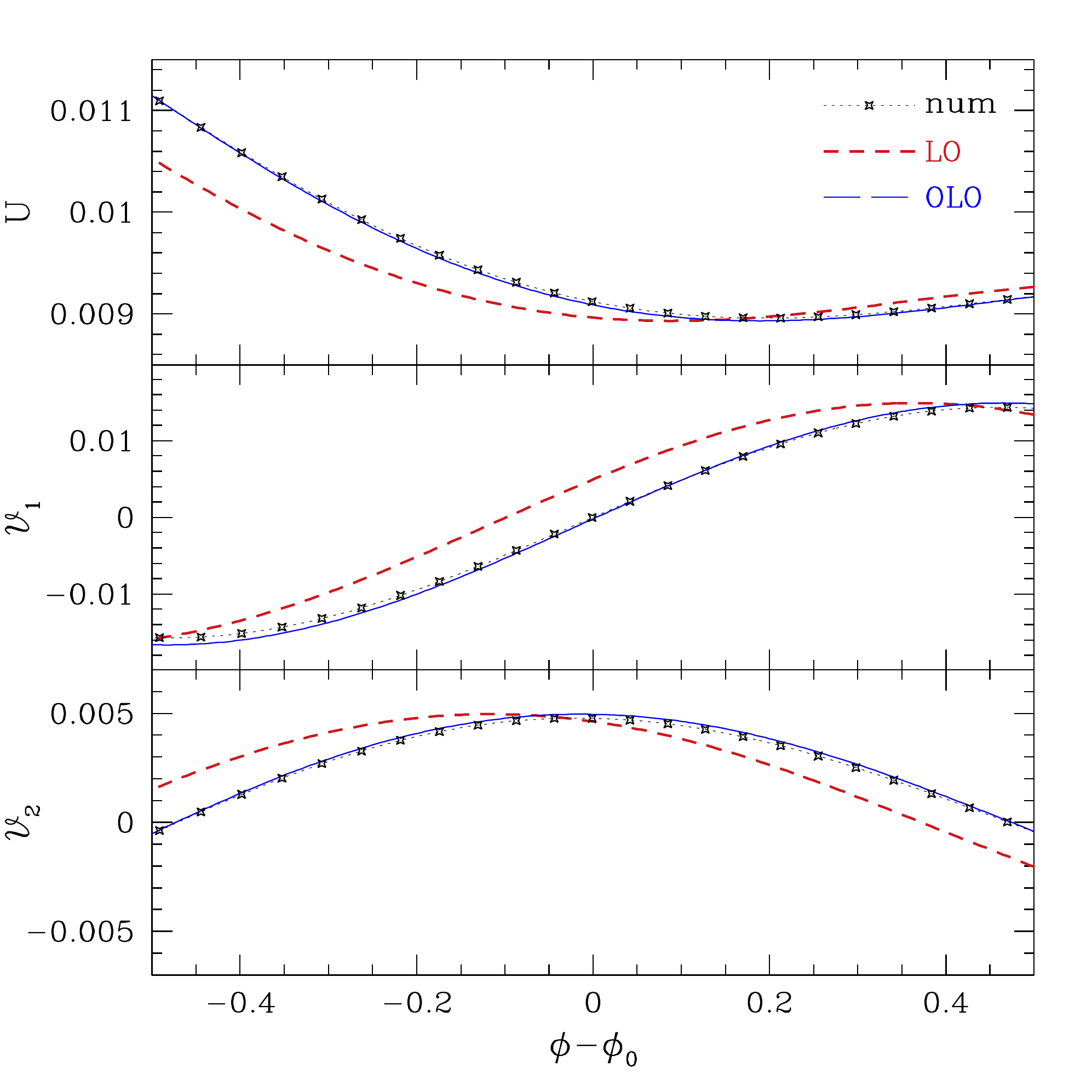}
\caption{Low frequency $\omega=1/3$ potential reconstruction in
  $U,\sV_1,\sV_2$ under the LO and OLO
  approximations as in Fig.~\ref{fig:DN3pow}.  Here the numerical results
  for $n_s,\alpha,r$ are used in each case.   The OLO approximation is 
  accurate at the $10^{-3}$ level whereas the LO approximation produces a systematic
  but consistent shift in field position.}
\label{fig:DN3pot}
\end{figure}

\begin{figure}[t]
\centering
\includegraphics[width=3in]{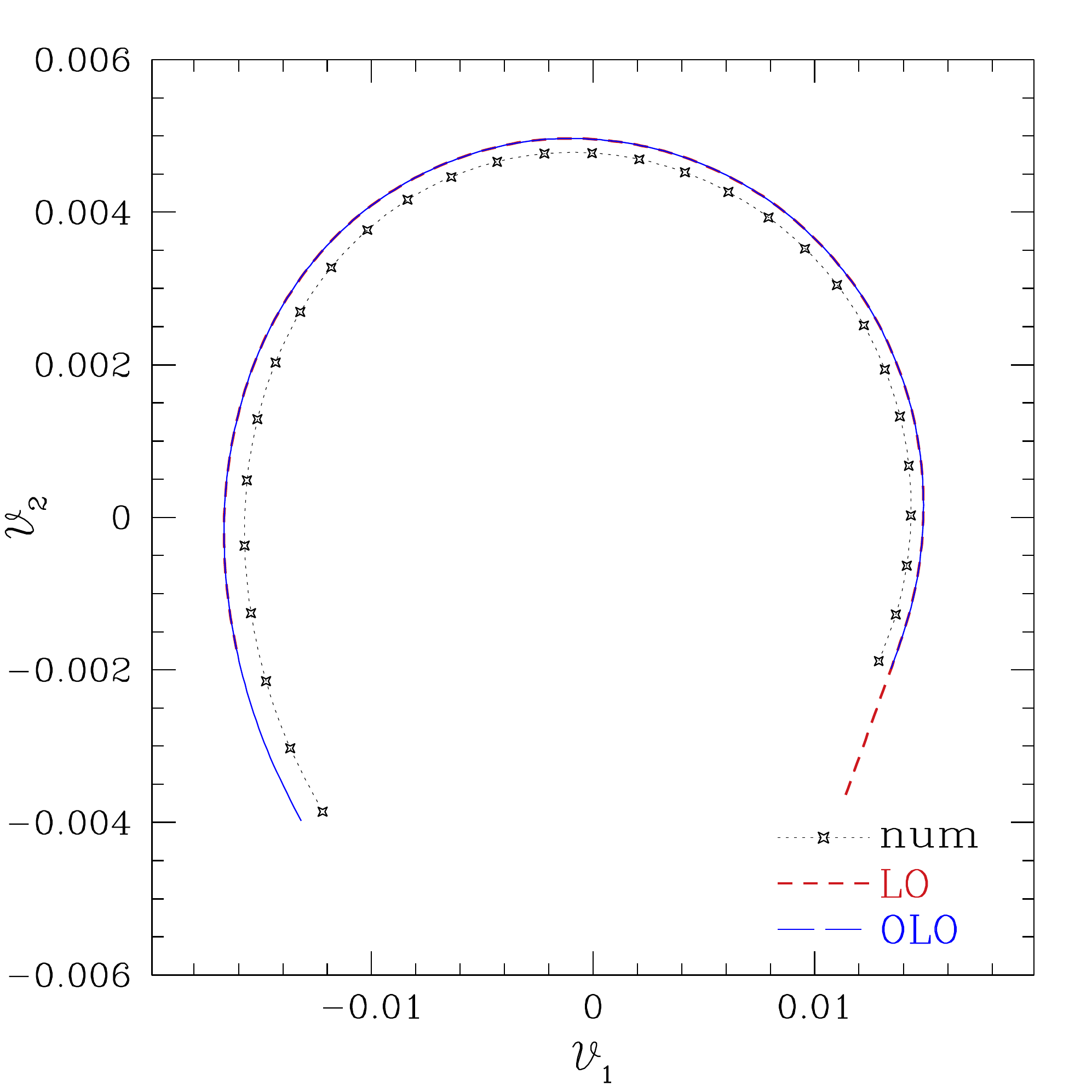}
\caption{Low frequency $\omega=1/3$ oscillation trajectories in the
  $\sV_1$-$\sV_2$ plane under the LO and OLO
  approximations as in Fig.~\ref{fig:DN3pow}.  Both LO and OLO trajectories
  are good approximations to the numerical results emphasizing that the LO results
  err simply in the evaluation point along the trajectory, parametrically evaluated here
  for the same range in observed wavenumber $k$.}
\label{fig:DN3V1V2}
\end{figure}

\begin{figure}[t]
\centering
\includegraphics[width=3.5in]{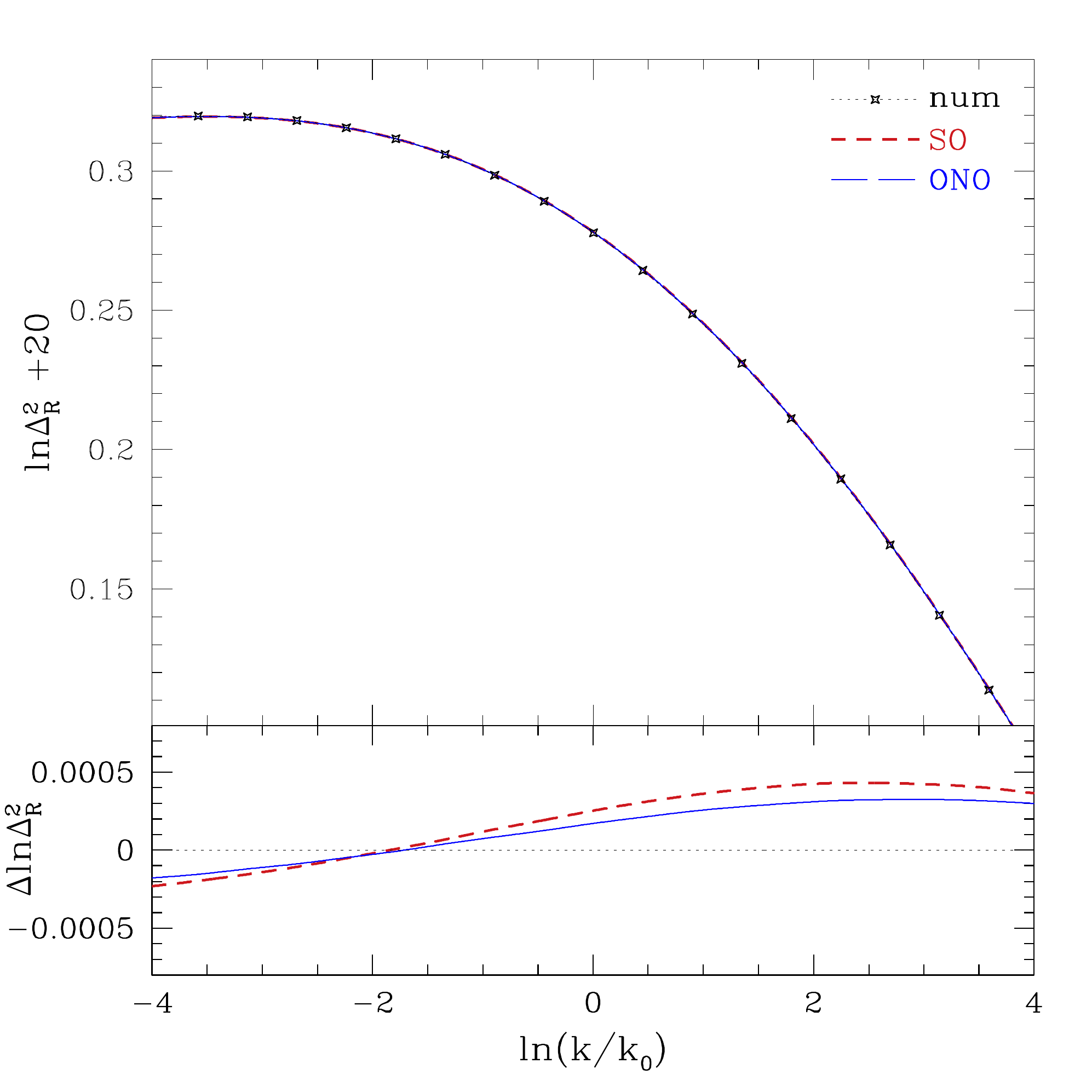}
\caption{Low frequency $\omega=1/3$ oscillations
  in the curvature power spectrum
  $\Delta_\curv^2$ under the standard second-order (SO) and optimized next-order (ONO) 
  approximations.  In both cases, the second-order corrections are sufficient to
  correct errors in Fig.~\ref{fig:DN3pow} to below the $10^{-3}$ level (bottom panel). }
\label{fig:DN3Spow}
\end{figure}

\begin{figure}[t]
\centering
\includegraphics[width=3.5in]{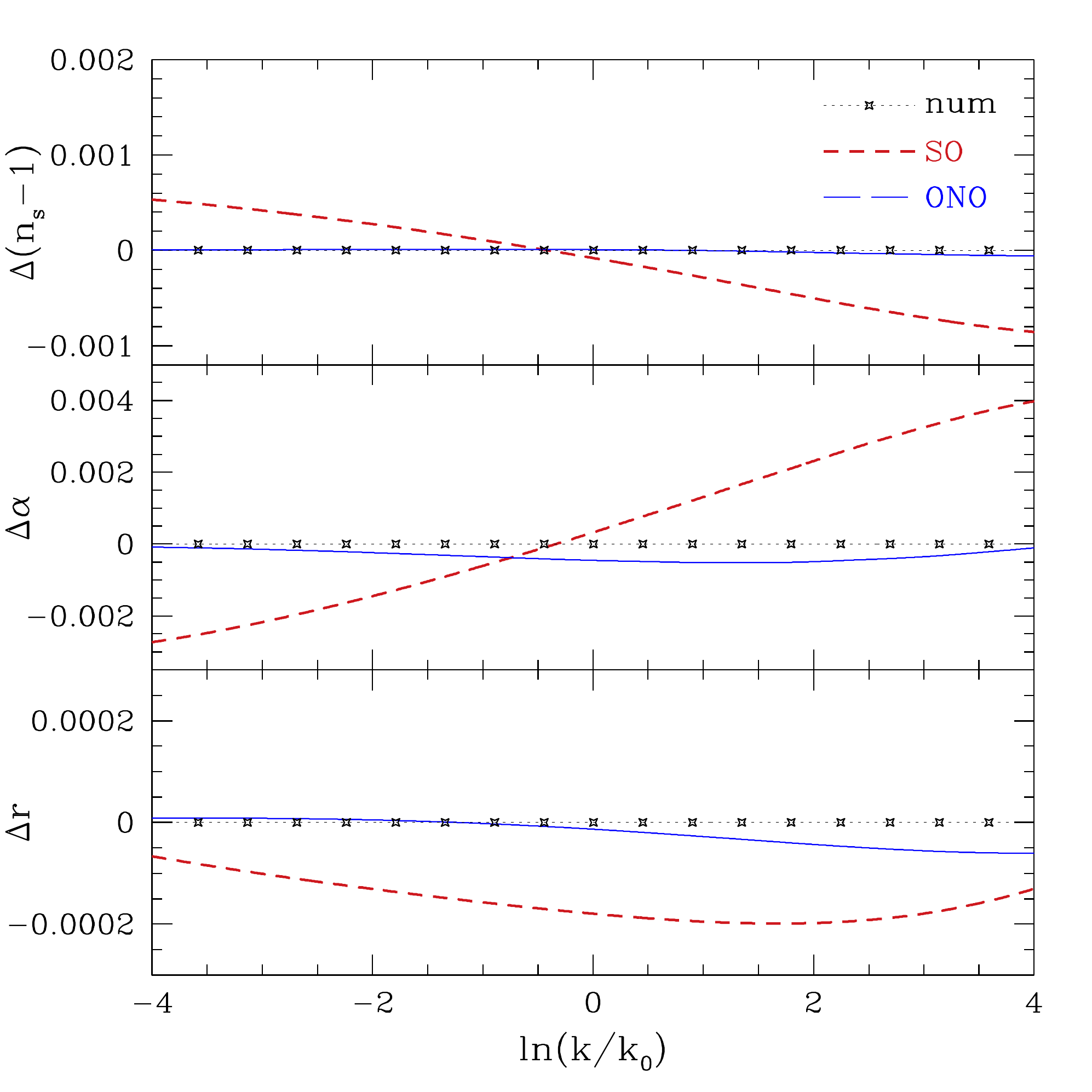}
\caption{Low frequency $\omega=1/3$ oscillations for the error in
  $n_s$, $\alpha$ and $r$  relative to numerics under the SO and ONO
  approximations.   ONO further improves OLO but  
  SO  only corrects LO in $n_s$ and $r$ compared with Fig.~\ref{fig:DN3obs}.  
  $\alpha$ remains shifted in SO and breaks the internal consistency of the
  observables. }
\label{fig:DN3Sobs}
\end{figure}

\begin{figure}[t]
\centering
\includegraphics[width=3in]{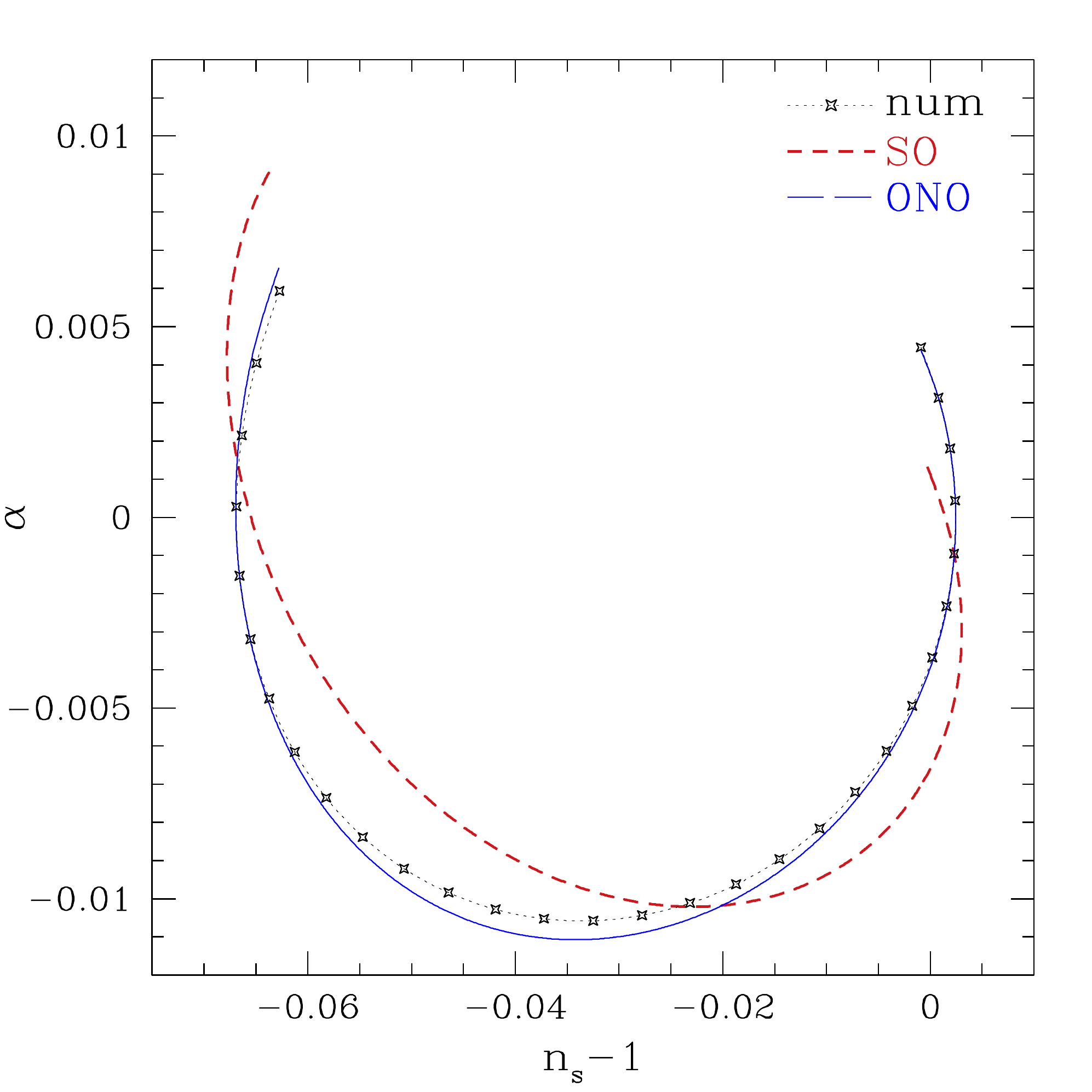}
\caption{Low frequency $\omega=1/3$ oscillation trajectories in the
  $n_s$-$\alpha$ plane under the SO and ONO. 
  ONO  improves OLO but SO is worse than the LO 
  (cf.~Fig.~\ref{fig:DN3nsalpha} and also Fig.~\ref{fig:DN3nsalphah} for the second-order
  Hubble flow result). }
\label{fig:DN3Snsalpha}
\end{figure}

By solving the exact Mukhanov-Sasaki equation (\ref{eqn:yeqn}) with Bunch-Davies initial 
conditions (\ref{eqn:BDcond}), we can compare
the power spectra observables to the various slow-roll approximations and
integral approaches.

We start by considering an example with $\omega=1/3$.    In Fig.~\ref{fig:DN3pow}, we
compare the standard LO and OLO  approximation of Eq.~(\ref{eqn:tiltrunVopt}) against
the numerical calculation for the curvature power spectrum $\Delta_\curv^2$.
The OLO approximation reproduces the numerical calculation at the $10^{-3}$ level,
while the LO approximation misestimates the power spectrum by a considerable
amount.   On the other hand, we know analytically that the two differ simply in that the power spectrum is
shifted by $\Delta \ln k \approx -\ln x_1$.  Since the relationship between $k$ and $\phi$
depends on the  physics of reheating, this type of error just introduces an
efold shift in those inferences rather than on the shape of the potential.

These considerations apply to $n_s$, $\alpha$, and $r$ as illustrated in Fig.~\ref{fig:DN3obs}.  
The OLO approximation is highly accurate with
differences from the numerics at the $10^{-3}$ level or less,
whereas the LO results are all shifted  by the same amount as 
the power spectrum.

This joint shift of all observables means that for the LO approximation all four
observables are consistent with arising from the same potential as would be inferred
from the better OLO approximation.   For example, we show trajectories in the $n_s$-$\alpha$ plane
in Fig.~\ref{fig:DN3nsalpha}.   The difference between the two approximations now
appears simply as a different starting and ending point on the same trajectory.

The importance of consistency becomes clear when comparing potential
reconstruction between the LO and OLO approximations in Eq.~(\ref{eqn:reconOLO}).
In Fig.~\ref{fig:DN3pot}, we show reconstruction using the exact, numerical
results for $n_s$, $\alpha$, and $r$ as a function of $\ln k$ as if they were precisely measured from data.   Again the OLO approximation is highly accurate 
for reconstruction, whereas the LO approximation shows a shift in $\phi$ due to the
change in the evaluation epoch

Consistency in the evaluation epoch between observables leads to consistency
in the potential reconstruction.   In Fig.~\ref{fig:DN3V1V2}, we show trajectories
in the $\sV_1$-$\sV_2$ plane.     The difference between the LO and OLO approximations
is again the starting and ending point on the trajectory.  Both approximations would
reconstruct potentials that were consistent with the true potential at the same level
of accuracy.  Though the values of $\sV_1$ and $\sV_2$ that they recover do differ, they
are consistent with a shifted evaluation on the potential.

\begin{figure}[t]
\centering
\includegraphics[width=3.5in]{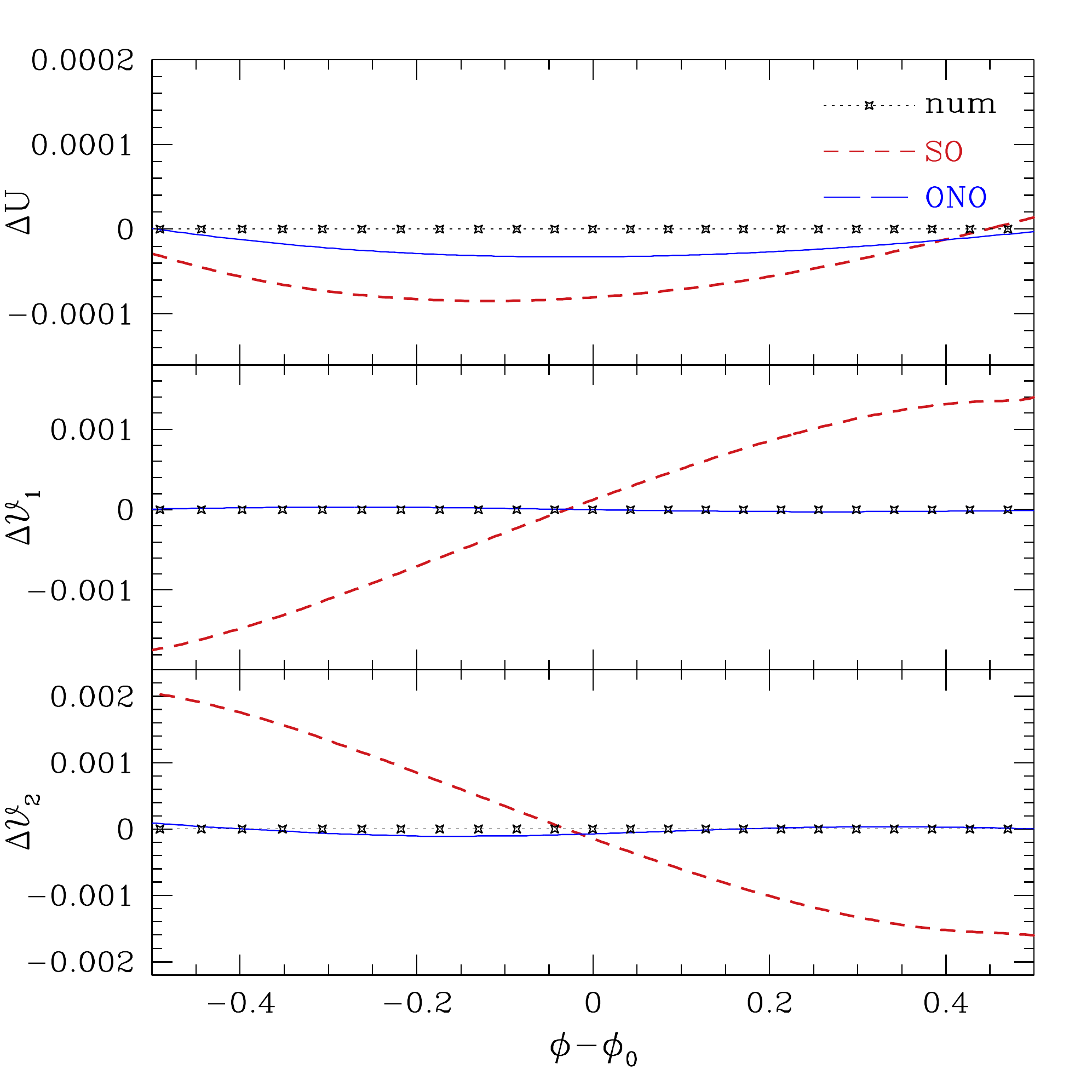}
\caption{Low frequency $\omega=1/3$ potential reconstruction error in
  $\U,\sV_1,\sV_2$ under the SO and ONO
  approximations.   ONO reconstructs the potential to extremely high accuracy ($\sim 10^{-4}$), but  
  SO corrects only $\U$ and $\sV_1$ compared with Fig.~\ref{fig:DN3pot} leaving
  $\sV_2$ with large fractional errors.  For ONO, we plot $\delta \tilde \sV_1$, $\delta \tilde \sV_2$. }
\label{fig:DN3Spot}
\end{figure}

\begin{figure}[t]
\centering
\includegraphics[width=3in]{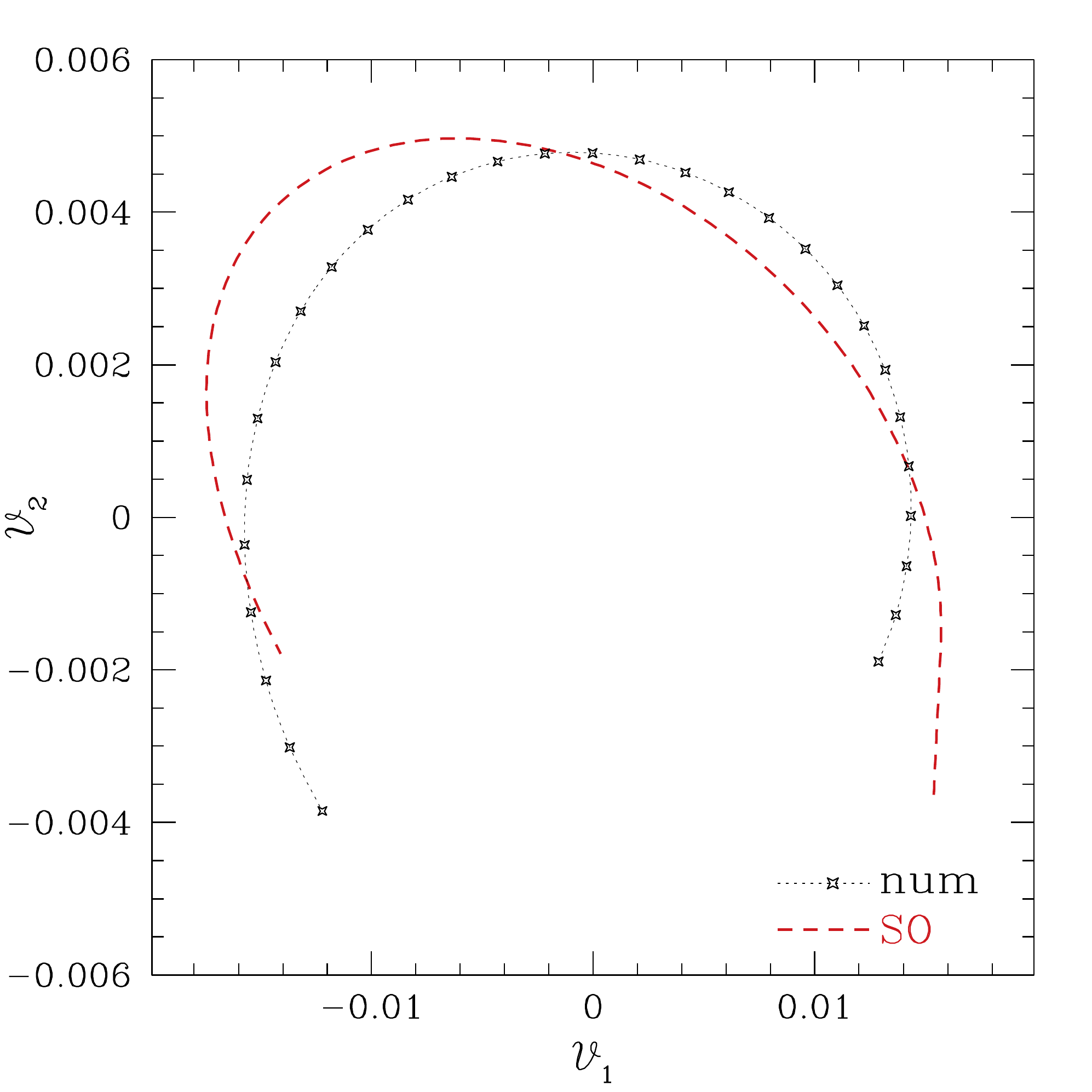}
\caption{Low frequency $\omega=1/3$ oscillation trajectories in the
  $\sV_1$-$\sV_2$ plane under the SO 
  approximation as in Fig.~\ref{fig:DN3pow}.   SO  is worse than  LO (cf.~Fig.~\ref{fig:DN3V1V2}) and can lead to an incorrect falsification of the true model.}
\label{fig:DN3SV1V2}
\end{figure}

\begin{figure}[t]
\centering
\includegraphics[width=3.5in]{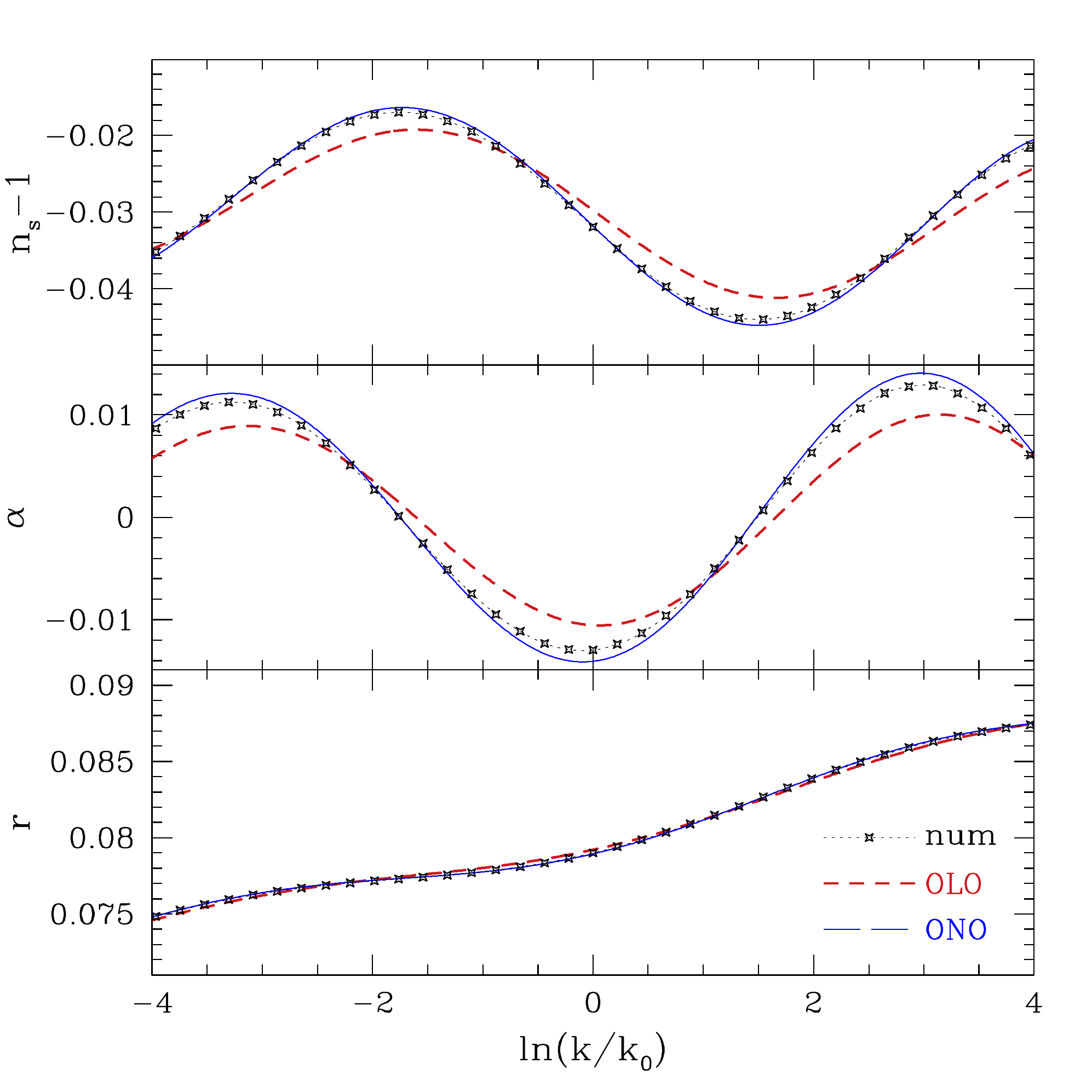}
\caption{Intermediate frequency $\omega=1$ oscillation in $n_s$, $\alpha$, and $r$  under the OLO and ONO
  approximations. OLO is still sufficiently accurate for current measurements, 
  but ONO provides notable corrections and preserves  $10^{-3}$ accuracy in observables. }
\label{fig:DN1Sobs}
\end{figure}

This should be contrasted with the standard second-order (SO) approximation.
In Fig.~\ref{fig:DN3Spow}, we show the SO and ONO approximations for $\Delta_\curv^2$ 
from Eqs.~(\ref{eqn:NO}) and (\ref{eqn:NOt}).
Both approximations do an excellent job of correcting residual errors in the LO and OLO approximations in the power spectrum itself.
For $n_s,\alpha,r$, Fig.~\ref{fig:DN3Sobs} shows that ONO  further improves the error
in their predictions from OLO to well below $10^{-3}$.  
Residual errors are largest for $\alpha$ due in large part to the neglect of the ${\cal O}(1/N^2\Delta N)$ term
$8 \tq_1 \U \sV_2$ in Eq.~(\ref{eqn:NO}) and would decrease for smaller choices of $\omega$.
On the other hand, SO only corrects
LO for $n_s-1$ and $r$ leaving the large fractional error in $\alpha$ that was present at LO.

Ironically this means that the SO approximation performs worse than LO in an 
important sense.   In Fig.~\ref{fig:DN3Snsalpha}, we highlight the problem for the
trajectories in the $n_s$-$\alpha$ plane.  Unlike the LO approximation, the SO 
approximation predicts pairs of these observables that are strongly inconsistent
with the true trajectory.

Likewise for potential reconstruction, the ONO approximation improves the reconstruction to
better than the $10^{-4}$ level.   In fact by defining fully self-consistent corrected potential
parameters $\tilde \sV_n$
from Eq.~(\ref{eqn:Vtilde}),  reconstruction is actually more accurate than the power spectrum prediction
since it absorbs  ${\cal O}(1/N^2\Delta N)$ terms into the definition.
On the other hand, while 
the SO approximation corrects the shift
of the LO approximation in $\U$ and $\sV_1$, it does not for $\sV_2$ as shown in
Fig.~\ref{fig:DN3Spot}.    The consequence is that in the $\sV_1$-$\sV_2$ 
plane shown in Fig.~\ref{fig:DN3SV1V2} the SO approximation 
reconstructs points that are not consistent with being anywhere on the
trajectory of the true potential.   This can lead to incorrect falsification of
the true model from observations.

Perhaps surprisingly, the OLO and ONO approximations still perform well for $\omega=1$
where all terms in the infinite $\sV_n$ series are of the same order.  This is related to the fact that the Taylor series still converges, albeit slowly, for even larger values up to  $\omega<2$.      In Fig.~\ref{fig:DN1Sobs}, we show their predictions for
$n_s-1$, $\alpha$ and $r$.   For the precision of current measurements, the OLO approximation actually still suffices.    For higher precision measurements, the ONO approximation now makes relatively
important corrections to the scalar observables and performs well at the $10^{-3}$ level.

\begin{figure}[t]
\centering
\includegraphics[width=3.5in]{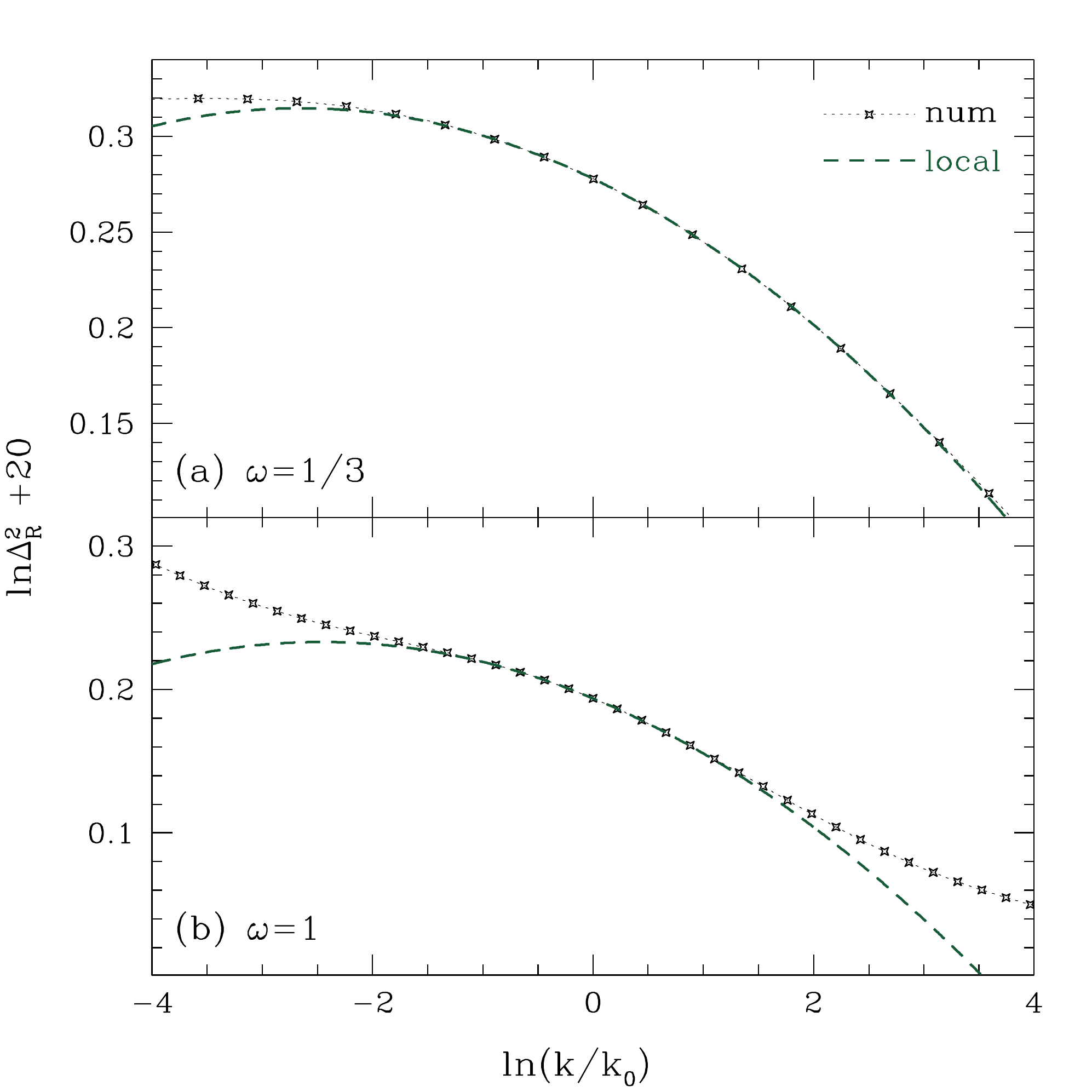}
\caption{Local, $n_s-\alpha$ parameterization of the power spectrum from
  Eq.~(\ref{eqn:local}) compared to numerics for the low and intermediate frequency cases. 
  Accuracy of the local approximation is directly related to the hierarchical structure of 
  the slow-roll parameters and hence that  of the OLO  truncation of the hierarchy. 
  For $\omega=1/3$ both suffice, whereas for $\omega=1$ both require correction from higher
  terms in their respective Taylor series.}
\label{fig:local}
\end{figure}

Finally, the accuracy of the OLO and ONO approximations are directly related to the
accuracy with which the local tilt and running describe the global power spectrum
across the observed efolds through Eq.~(\ref{eqn:local}).  In Fig.~\ref{fig:local}, we show that
in the $\omega=1/3$ case where the OLO approximation suffices this approximation is very accurate over a large range in efolds.    For the $\omega=1$ case, the approximation only holds for the central $\sim 4$ efolds and likewise the ONO corrections to OLO become important.   
Thus, if the observations are a good fit to the local model, then OLO is a good
approximation, and ONO provides the means to improve its accuracy if necessary in the future.   
If the local model is a bad fit to the data, then $\Delta N \lesssim 1/2$ (or $\omega \gtrsim 2$), 
and no slow-roll hierarchy of parameters can describe the spectrum accurately.

\subsection{High Frequency Oscillations}
\label{sec:highfrequency}

For $\omega > 2$, the slow-roll hierarchy does not converge with a finite number of terms. 
Nonetheless, the GSR expansion  remains valid so long as the deviations from scale invariance, controlled by the amplitude not the frequency of the oscillation, remains small.
We must, however, evaluate the GSR integrals in Eqs.~(\ref{eqn:GSR}) and (\ref{eqn:I1I2}) directly.
Given that in the high frequency limit
$\sV_p$ increases with $p$, we can no longer  naively use their relationships with
$G^{(p)}$ given in the Appendix.
Reference~\cite{Dvorkin:2009ne} showed that, even in the high frequency limit,
\begin{equation}
G' = 3\U - 2\sV_1 + \epsilon_H {\cal O}(G') + {\cal O}(\delta_1^2).
\label{eqn:Gprimelarge}
\end{equation}
We can therefore establish the leading-order connection to the potential in the usual way.
Namely for the oscillatory part,
\begin{equation}
\delta G = 3 \frac{\delta V}{\bar V} - 2 \frac{\delta V^{(1)}}{\bar V^{(1)}} .
\label{eqn:dGm}
\end{equation}
Moreover, this leading-order result remains valid  even
for large amplitude, high frequency oscillations since $G'$ is linear in the highest
derivative term $V^{(2)}$ (or $\delta_2$) and hence linear in $\Lambda^4$ in that limit.    This extended validity 
will be important in the consideration of quadratic terms below. It can be demonstrated
by integrating $\delta G'$ assuming that the field rolls according to the smooth potential.
Even large amplitude high frequency oscillations  give highly suppressed effects on the field position which we can neglect.

Under this approximation for the roll, 
we can convert field position to time relative to some fiducial evaluation point
$\eta_*$ as
\begin{equation}
\phi \approx \phi_* + \sqrt{2\bar \epsilon_H} \ln (\eta/\eta_*)  \approx \phi_* + \frac{1}{\phi_*}\ln (x/k\eta_*)
\label{eqn:phisr}
\end{equation}
to obtain the leading-order contribution from the oscillations
$\delta \ln \Delta_\curv^2(k) \approx \delta I_0(k)$,
where
\begin{eqnarray}
\delta I_0  
&\approx&  -\frac{2\Lambda^4}{ \lambda f } \int_{0}^{\infty} \frac{dx}{x}  W'(x)  \Bigg[ \sin\left( \omega_* {\ln x}  + \psi\right) \nonumber\\
&&
+\frac{3 f}{2 \phi_*}  \cos\left( \omega_* {\ln x}  + \psi\right)\Bigg],
\end{eqnarray}
with $\omega_* = (f \phi_*)^{-1}$ and 
\begin{equation}
\psi (k) = \frac{\phi_*}{f}  - \omega_*  \ln(k \eta_*)   +\theta.
\label{eqn:psiphase}
\end{equation}
As in the optimized slow-roll calculations, 
the epoch $\eta_*$ and the associated field value $\phi_*$ can be  optimized to make the approximations as accurate as possible for
a given $k$.    It does not have to coincide with the fixed $\eta_0$ and $\phi_0$,
associated with horizon crossing for $k_0$.

We can evaluate both terms in the integral using Eq.~(\ref{eqn:intf}),
\begin{eqnarray}
&&\int_0^\infty \frac{dx}{x}  W'(x) e^{i \omega_* \ln x } \nonumber\\
&& \qquad =  - 2^{-i\omega_*} \cosh \mk{\f{\pi \omega_*}{2}} \f{3\Gamma(2+i\omega_*)}{(1-i\omega_*)(3-i\omega_*)} \nonumber\\
&& \qquad  = \sqrt{ \frac{9 \pi \omega_* \coth(\pi \omega_* /2)}{ 2(9+ \omega_*^2)} } e^{i (\pi/2-\beta)}.
\end{eqnarray}
The phase factor $\beta(\omega_*)$ can be expressed exactly in terms of the $\Gamma$
function formula.   
Note that it takes on simple limiting forms
\begin{eqnarray}
\lim_{\omega_* \rightarrow 0} \beta &=& \frac{3\pi}{2} -\omega_* \ln x_1 ,\nonumber\\
\beta_\infty \equiv \lim_{\omega_* \rightarrow\infty} \beta &=& \omega_*[1-\ln(\omega_*/2)] -\frac{\pi}{4}.
\label{eqn:beta}
\end{eqnarray}
 Here $\ln x_1\approx 1.06$ is the same factor that enters into the OLO approximation for reasons that will
be clear below.   
The large $\omega_*$ limit corresponds to the resonant or stationary phase solution of
$x=\omega_*/2$ for the
phase
factor $2x - \omega_*\ln x$ \cite{Flauger:2009ab}.  To clarify the connection between these
limits, it is useful to have a simple approximation to the exact $\beta(\omega_*)$
and we find that
\begin{equation}
\beta(\omega_*) \approx \beta_\infty+ \frac{7\pi}{4} \frac{1}{\sqrt{1+ 1.59\omega_*^{0.92} + \omega_*^2}}
\end{equation}
to better than 0.03 accuracy everywhere.

Putting these results together, we obtain
\begin{eqnarray}
\delta I_0(k) & =&  -A \left[ \cos(\psi-\beta)- \frac{3 f}{2 \phi_*} \sin(\psi -\beta)\right] ,
\label{eqn:monodromyI0}
\end{eqnarray}
where
\begin{eqnarray}
A  = \frac{12 \Lambda^4}{ \lambda f \sqrt{ 1+ (3 f\phi_*)^2}} \sqrt{\frac{\pi}{8}f\phi_* \coth\mk{\frac{\pi}{2 f\phi_*}} }.
\end{eqnarray}

This should be compared to the result from Ref.~\cite{Flauger:2009ab},
\begin{eqnarray}
\delta  \Delta^2_\curv &=& \bar A_s \left(\frac{k}{k_0} \right)^{\bar n_s-1}  A
\cos\left[  \omega \ln(k/k_0) +\varphi \right] ,
\label{eqn:Flauger}
\end{eqnarray}
where $\varphi$ is considered a free phase parameter.    
Since this form is primarily used for the purpose of fitting a functional form for the deviations to data, in this comparison
differences in the form are more important than those in the association of $A$, $\omega$, and
$\varphi$ with parameters in the potential.  On the other hand, our treatment has the advantage of preserving all of those relations.

For example, 
Ref.~\cite{Flauger:2009ab} ignores corrections of order $ f/\phi_* \approx {2\epsilon_H}/\omega_*$ as they are not relevant for high frequency oscillations, 
but these may be absorbed into
a redefinition of amplitude and phase.   More importantly they take
\begin{equation}
\delta \ln \Delta_\curv^2 \approx \frac{ \delta \Delta_\curv^2}{\bar\Delta_\curv^2}
\end{equation}
which is valid in the $|A| \ll 1$ limit but causes differences that cannot be reabsorbed into
fits at large $A$.  Note that Eq.~(\ref{eqn:Flauger}) does not guarantee a positive definite
power spectrum.

In terms of the phase, Eq.~(\ref{eqn:Flauger}) defines only the $k$
dependence rather than the absolute relationship between $\varphi$ and $\theta$ 
given by our treatment.  The dependence of the phase on $k$  in Eq.~(\ref{eqn:psiphase}) 
is consistent with Eq.~(\ref{eqn:Flauger})
if we choose the evaluation epoch as $\eta_*=1/k_0$,
\begin{equation}
\psi \approx \frac{\phi_*}{f} - \omega_* \ln (k/k_0) +\theta.
\end{equation} 
The advantage of the more general Eq.~(\ref{eqn:monodromyI0}) is that we are not required
to do so.  Using a fixed evaluation epoch omits the evolution of quantities like
$\bar\epsilon_H$ between that epoch and the true freeze-out epoch which depends on $k$.
In the more refined treatment of Ref.~\cite{Flauger:2014ana}, these effects are absorbed into a running or
drift of the frequency.  Note that this type of running does not invalidate our use of the slow-roll approximation in
Eq.~(\ref{eqn:phisr}) which only requires constant $\bar \epsilon_H$ during freeze-out of a given $k$ rather than across
many efolds of $k$ and hence $N$ space.

We can instead  parallel and extend our low frequency treatment by allowing the freeze-out epoch for the oscillatory piece $\delta V$  to 
depend on $k$ in potentially a different way than for $\bar V$.  We start by reexamining
the low frequency case where the two freeze-out epochs should be the same.   Indeed, if we take the evaluation point as $k\eta_* = x_1$,
the $\ln x_1$ phase shift in $\beta$ in Eq.~(\ref{eqn:beta}) cancels with the opposite shift in $\psi$ in Eq.~(\ref{eqn:psiphase}) so that
\begin{equation}
\lim_{\omega_* \rightarrow 0}( \psi -\beta )=  \frac{\phi_*}{f}  +\theta  + \frac{\pi}{2}, \quad k \eta_*  = x_1.
\end{equation}
Equation~(\ref{eqn:monodromyI0}) in the $\omega \ll 1$ limit is then exactly the  OLO approximation
using Eq.~(\ref{eqn:dGm}):
\begin{eqnarray}
\delta I_0 & \approx  &\delta G|_{x=x_1}, \quad
({\rm OLO}) \\
&\approx& \frac{2\Lambda^4}{\lambda f}  \Bigg[ \sin\left( \frac{\phi_*}{f} +\theta \right) 
+\frac{ 3f}{2\phi_*}  \cos\left( \frac{\phi_*}{f} +\theta \right)\Bigg]. \nonumber
\end{eqnarray}
Note that here
$\phi_*$ is a function of $k$ due to the optimized evaluation, and the explicit dependence can either be directly computed given
the background solution or characterized by the running of the frequency from some central value at $k_0$
\cite{Flauger:2014ana}.

For high frequency oscillations $\omega \gg 1$, we increase accuracy by shifting the evaluation point $\eta_*$ to the stationary phase point
\begin{equation}
\lim_{\omega_*\rightarrow \infty}( \psi -\beta )= \frac{\phi_*}{f}  -\omega_*  +\theta + \frac{\pi}{4}, \quad k  \eta_* = \frac{\omega_*}{2}.  
\end{equation}
Optimized evaluation again means that $\phi_*(k)$ is given by the solution to
\begin{equation}
k\eta(\phi_*) = \frac{\omega_*}{2} = \frac{1}{2 f \phi_*}.
\end{equation}
To cover both low and high frequency cases, we can combine the two 
freeze-out criteria as
\begin{eqnarray}
k \eta_* &=& {\rm max} \mk{x_1, \frac{\omega_*}{2}} .
\end{eqnarray}

\begin{figure}[t]
\centering
\includegraphics[width=3.5in]{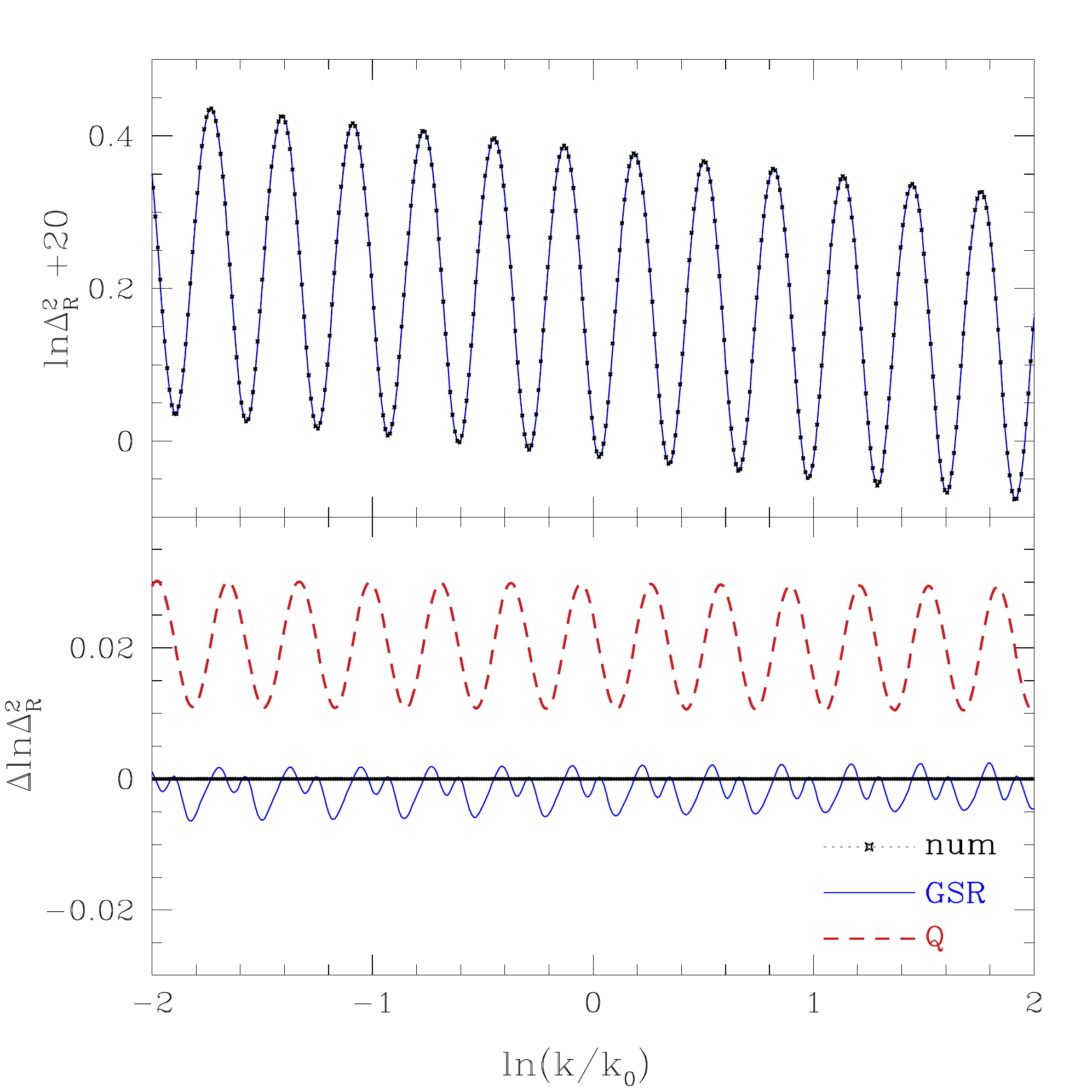}
\caption{High frequency $\omega=20$ oscillation in the curvature power spectrum
  $\Delta_\curv^2$ under the analytic GSR integral approximation as compared with numerical
  calculations for a model with $A \sim 0.2$ amplitude oscillations (top).   Differences
  are less than $10^{-2}$ with a significant correction coming from the quadratic terms 
  $(Q)$ which mainly change the phase and amplitude of the oscillation as well as shift
  the normalization parameter $\ln \bar A_s$ (bottom). }
\label{fig:highfreq}
\end{figure}

For the second-order corrections in the $\omega \ll 1$ limit, the ONO approximation applies and captures all  terms.
In the opposite limit of $\omega \gg 1$, quadratic terms in $A$ can be more relevant since a large amplitude 
oscillation does not produce large changes to the well-constrained average tilt and running.  In this
large amplitude and frequency limit,
the dominant second-order correction is due
to $I_1$ \cite{Dvorkin:2009ne}
where a similar resonance occurs in the stationary phase approximation.
Thus, the oscillatory part of the power spectrum is well approximated by
\begin{equation}
\delta \ln  \Delta_\curv^2 \approx \delta I_0 + 2\bar I_1 \delta I_1 + \delta I_1^2, \quad ({\rm GSR})
\label{eqn:dlnD2}
\end{equation}
where the smooth and oscillatory parts of $I_1$ are given by
\begin{equation}
\bar I_1= \frac{\pi}{2\sqrt{2}}(1-\bar n_s),
\end{equation}
and
\begin{eqnarray}
\delta I_1 &=& { 1\over \sqrt{2} } \int_0^\infty {d x \over x} \delta G'(\ln x) X(x) .
\end{eqnarray}

Repeating the same steps as for $I_0$, we have integrals that can be evaluated using
\begin{eqnarray}
&&\int_0^\infty \frac{dx}{x}  X'(x) e^{i \omega_*\ln x} \nonumber\\
&&\qquad= - 2^{-i\omega_*}i  \sinh \mk{\f{\pi \omega_*}{2}} \f{3\Gamma(2+i\omega_*)}{(1-i\omega_*)(3-i\omega_*)} \nonumber\\
&&\qquad = - \sqrt{ \frac{9 \omega_*\pi \tanh(\pi \omega_* /2)}{ 2(9+ \omega_*^2)} } e^{-i\beta},
\end{eqnarray}
to obtain
\begin{equation}
\delta I_1 = \frac{A}{\sqrt{2}} \tanh \mk{\f{\pi \omega_*}{2}} 
\left[ \sin(\psi-\beta) + \frac{3 f}{2 \phi_*} \cos(\psi-\beta)\right] .
\end{equation}
Note that the  $\delta I_1^2$ correction contains squared or $2\omega_*$ frequency contributions that cannot fully be absorbed into the form of Eq.~(\ref{eqn:Flauger}).  However to quadratic order in $A$, the combination of the $I_0$ and $I_1$
differences  in form between Eqs.~(\ref{eqn:dlnD2}) and (\ref{eqn:Flauger}) can be absorbed into a redefinition of $\bar A_s$, $A$ and the oscillation phase
for $\omega\gg 1$.
To see this note that, aside from the phase $\psi-\beta$, these quadratic differences are given by
\begin{equation}
Q=\frac{\delta I_0^2}{2}+ 2\bar I_1 \delta I_1 + \delta I_1^2
\end{equation}
and $\sqrt{2} \delta I_1$ and $\delta I_0$ are related by a phase shift of $\pi/2$ such that their squares sum to a constant since $\tanh(\pi\omega_*/2)\rightarrow 1$.

As an example, we choose $\omega=20$ and  $\Lambda^4=1.78\times 10^{-13}$ which gives
$A(k_0) \approx 0.2$.  Other parameters are set as in the low frequency case.  We use the ONO approximation to determine $\bar A_s$, $\bar n_s$, and $\bar \alpha$ 
for the smooth potential and add its contribution using 
Eq.~(\ref{eqn:local}) to compare the total
\begin{equation}
\ln \Delta_\curv^2 =  \ln \bar\Delta_\curv^2 \Big|_{\rm ONO} + \delta \ln \Delta_\curv^2 
\end{equation}
to numerical results.
Note that, by implementing the ONO approximation, where $\bar V$ is evaluated at
$k \eta(\phi_2)=x_2$, we use two field positions $\phi_*$ and $\phi_2$ for
each $k$ which allows us to simultaneously optimize for the oscillatory and smooth parts of the potential.

In Fig.~\ref{fig:highfreq}, we show that even for this large amplitude oscillation, Eq.~(\ref{eqn:dlnD2}) is accurate to better than $10^{-2}$ without any adjustment to the phase, frequency or amplitude of oscillations.
Note that, as expected, the quadratic terms in  $Q$, while significant in establishing the good agreement
with numerics, takes the form of a constant plus an oscillatory piece of the same frequency
as the leading-order contribution.
Thus, it can be absorbed into a redefinition of parameters in Eq.~(\ref{eqn:Flauger}) 
and the dependence of the oscillation phase $\psi-\beta$ on $k$.
For even larger values of the amplitude $A$, 
this is no longer possible and so accurate expressions will require
calculations to cubic and higher order in $A$

\section{Discussion}
\label{sec:discussion}

By utilizing the GSR approximation,  we  have provided a systematic study 
of the evaluation and interpretation of
power spectra where a relatively large local running arises from features in the potential.
This approach assumes only that the average deviation from scale
invariance is small  and associated with the number of efolds to the end of inflation as $1/N$.   
The frequency of temporal features $1/\Delta N$ can be much larger than $1/N$.   
We introduced the GSR slow-roll hierarchy parameters $G^{(p)}$ 
which are directly related to observables and elucidate their relation to the
standard hierarchies of parameters in the potential and Hubble flow.
This parameterization works for $P(X,\phi)$ models as well with 
a suitable generalization of the conformal time to the sound horizon.

For models with $1 < \Delta N \ll N$, a slow-roll hierarchy of parameters still exists,
but the running of the tilt $\alpha$ is only smaller than $n_s-1$ by a factor of $\Delta N$ not $N$.
In these models, the leading-order slow-roll calculation 
gains relatively large corrections since a large
running implies that the slow-roll parameters are not constant.
Instead the calculation and interpretation of such models 
proceeds by Taylor expanding the temporal evolution.
This series converges rapidly as long as $\Delta N \gtrsim$ few, 
and this condition also guarantees that observable power spectra are well characterized
by the local tilt and running across the well-measured few efolds of the CMB 
and large scale structure.

In fact next-to-leading-order corrections in the Taylor series of each observable
can be consistently reabsorbed into an optimized temporal evaluation of leading-order terms.
This is because the first Taylor correction just represents a shift in the epoch of fluctuation
freeze-out to about an efold before horizon crossing.
The advantage of this OLO approach is that the analysis of observables proceeds exactly in
the same way as a standard leading-order analysis and only the interpretation in terms
of their correspondence to the inflationary model differs.   For the interpretation of
current data fits to constant tilt and running of the tilt, this simple approach suffices in accuracy.
In terms of potential reconstruction it recovers a local cubic
expansion evaluated consistently at the optimized field point.

Contrast this with a common approach in the literature that attempts to correct for
the evolution with a second-order expansion under the assumption that $\Delta N \sim N$.
In this case the tilt is corrected for evolution between horizon crossing and freeze-out 
but the running of the tilt, which is assumed to be intrinsically second order, is not.
In this case, the observables are effectively evaluated at inconsistent epochs.   
Potential reconstruction from a second-order
approach likewise produces inconsistencies  which
could potentially lead to an incorrect falsification of the true model from the observations.
For models with large running
it can only be consistently applied to a purely, rather than locally, cubic potential.
Ironically then the second-order analysis provides a more complicated but less
general approach compared with the leading-order analysis as we explicitly demonstrate
using a low frequency axion monodromy model.

For even higher accuracy, we can keep both 
$1/N^2$ terms and  optimized next-to-leading-order terms in the hierarchy.   This ONO
approach leads to better than $10^{-3}$ accuracy in
observables and potential reconstruction even for
$\Delta N\sim 1$ where all of the infinite hierarchy of parameters have the same magnitude.

Finally for $\Delta N \ll 1$, the slow-roll parameters possess an inverted hierarchy and lose
their utility for predicting observables.    Even in this case the GSR approach provides
accurate predictions as long as the amplitude of deviations from time translation invariance is
small regardless of the frequency of temporal features. 
Likewise the technique of optimizing the epoch of evaluation can be generalized to
establish precise relationships between the potential parameters and observables.

We test this GSR approach with
axion monodromy models in the high frequency limit and provide expressions that 
are accurate to second order in the amplitude and optimized in the temporal evaluation.   
The optimized approach also provides the direct
relationship between parameters of the potential and phenomenological template fits
to the running phase, frequency and amplitude of oscillations in the power spectrum 
used in the literature.   
For sufficiently large amplitude deviations and/or future high precision measurements, 
relating power spectrum features to potential parameters requires
going beyond the second order calculations presented here.
We leave such considerations and their impact on interpreting observables to a future work.

\acknowledgments
We thank A.\ Joyce for useful discussions and C.\ He and V.\ Miranda for cross-checking of numerical results. 
H.M.\ was supported in part by Japan Society for the Promotion of Science 
Postdoctoral Fellowships for Research Abroad.
W.H.\ was supported by NASA ATP NNX15AK22G, US Department of Energy Contract No.\ DE-FG02-13ER41958 
and the Kavli Institute for Cosmological Physics at the University of
Chicago through Grants No.\ NSF PHY-0114422 and No.\ NSF PHY-0551142.

\appendix

\section{Parameter Relations}

To relate the three different parameterizations of the slow-roll hierarchy
$G^{(p)}$, $\{\epsilon_H,\delta_p \}$ and $\{ \U, \sV_p\}$ used in the main text,
we establish here the relationship between the GSR, Hubble slow-roll, and potential
slow-roll approaches.

We expand expressions to ${\cal O}(1/N^2)$ under the assumption that
\begin{eqnarray}
\{ \epsilon_H,\U \} &=& {\cal O}\left(\frac{1}{N}\right) ,\nonumber\\
\{ G^{(p)},\delta_p,\sV_p\} &=& {\cal O}\left(\frac{1}{N\Delta N^{p-1}}\right).
\end{eqnarray}
with $1 \ll \Delta N \le |N|$.
Unlike the standard slow-roll approximation, we therefore
keep terms that are  ${\cal O}(1/N \Delta N^p)$ but still drop those that
are $O(1/N^2 \Delta N^p)$ for $p\ge 1$.   For example we keep $\sV_2$ and $\U \sV_1$ but
omit $\U \sV_2$.

The relationship between the GSR and Hubble slow-roll variables involves
the conversion between conformal time and efold 
\begin{equation}
\frac{dN}{d\ln \eta}=  - aH \eta.
\end{equation}
This quantity can be parameterized by using Eqs.~(\ref{eqn:eps}) and (\ref{eqn:hubbleeom})
to obtain
\begin{eqnarray} 
\frac{d H }{dN} & =& -\epsilon_H H,\nonumber\\
\frac{d^2 H}{dN^2} &=& -\epsilon_H(\epsilon_H+ 2 \delta_1) H .
\end{eqnarray}
Integrating the Taylor expansion of $H$ around $N$ to find $\eta= \int_N^0 dN/e^N H$,
we obtain
\begin{equation}
 aH \eta \approx 1 +\epsilon_H + 3\epsilon_H^2 + 2 \epsilon_H\delta_1.
\end{equation}
Using this relation, the definition of $f$ from Eq.~(\ref{eqn:fdef}), and the Hubble slow-roll 
hierarchy equation (\ref{eqn:hubbleeom}), we  obtain for the GSR scalar  parameters
 \begin{eqnarray}
\ln f^2 & \approx& \ln \left( \frac{8\pi^2\epsilon_H }{H^2} \right) + 2 \epsilon_H + 5\epsilon_H^2 + 4 \epsilon_H \delta_1, \nonumber\\
\frac{f'}{f}  &\approx& -\delta_1 - 2 \epsilon_H - 4 \epsilon_H^2 - 3\epsilon_H \delta_1,\nonumber \\
G' &\approx&  4\epsilon_H +2 \delta_1 + \frac{2}{3} \delta_2 + \frac{32}{3}\epsilon_H^2 + \frac{28}{3}\epsilon_H \delta_1 - \frac{2}{3}\delta_1^2, \nonumber\\
G'' & \approx & -2 \delta_2 -\frac{2}{3} \delta_3  -8 \epsilon_H^2  -10\epsilon_H\delta_1 + 2\delta_1^2, \nonumber\\
G^{(p)} & \approx & (-1)^{p+1} \left( 2 \delta_p+ \frac{2}{3} \delta_{p+1} \right),\qquad (p>2)
\label{eqn:secondsource}
\end{eqnarray}
and for the tensor parameters
\begin{eqnarray}
\ln f_h^2 &\approx& \ln \left( \frac{2\pi^2}{H^2} \right) + 2\epsilon_H + 5\epsilon_H^2 +4 \epsilon_H\delta_1, \nonumber\\
\frac{f'_h}{f_h} &\approx& -\epsilon_H - 3\epsilon_H^2 - 2\epsilon_H \delta_1,\nonumber \\
G'_h &\approx& 2 \epsilon_H + \frac{22}{3}\epsilon_H^2 + \frac{16}{3}\epsilon_H \delta_1,\nonumber\\
G''_h &\approx& -4\epsilon_H^2 - 4 \epsilon_H\delta_1, \nonumber\\
G_h^{(p)} &\approx & 0 ,\qquad (p>2).
\end{eqnarray}

For the relationship with the potential parameters,
we iteratively solve Eq.~(\ref{eqn:U})\begin{eqnarray}
\U &=& 2\epsilon_H + {\cal O}(N^{-2}) ,\nonumber\\
\sV_1 &=& \epsilon_H - \delta_1 -\f{\delta_2}{3} + {\cal O}(N^{-2})
\end{eqnarray}
along with
\begin{eqnarray}
\sV_p &=& -\frac{d \sV_{p-1}}{dN} +  {\cal O}(N^{-2}) \nonumber\\
&=& (-1)^p \mk{ \delta_{p}+ \f{\delta_{p+1}}{3} } + {\cal O}(N^{-2}) 
\end{eqnarray}
for $p\ge 2$.
Notice that this is the same combination that enters into $G^{(p)}$ in Eq.~(\ref{eqn:secondsource}).
Inverting these relations we obtain
\begin{eqnarray}
\epsilon_H  &=& \frac{\U}{2} +  {\cal O}(N^{-2}),  \nonumber\\
\delta_1 &=& \frac{\U}{2} - \sum_{n=0}^{\infty} \left(\frac{1}{3}\right)^n  \sV_{n+1} +  {\cal O}(N^{-2})  , \nonumber\\
\delta_p &=& \sum_{n=0}^{\infty} \left(\frac{1}{3}\right)^n  \sV_{n+2}+  {\cal O}(N^{-2})  .
\end{eqnarray}
We then plug these back into the exact Eqs.~(\ref{eqn:background}) and (\ref{eqn:U}) and use
\begin{eqnarray}
\frac{V'}{V} \frac{ d \U}{d\phi} &=& 2 \U (\sV_1-\U), \nonumber\\
\frac{V'}{V} \frac{ d \sV_p}{d\phi} &=& \sV_{p+1} + [(p-1) \sV_1 - p\U] \sV_p,
\end{eqnarray}
to obtain
\begin{eqnarray}
\frac{3 H^2}{V}  &\approx&  1+ \frac{ \U}{6} - \frac{\U^2}{12}  + \frac{\U \sV_1}{9}  ,
\nonumber\\
\frac{\epsilon_H}{\U}  &\approx &  \frac{1}{2} - \frac{\U}{3}+ \sum_{n=1}^{\infty} \left(\frac{1}{3}\right)^n  \sV_n   +  \frac{4}{9} \U^2 - \frac{5}{6} \U \sV_1 +  \frac{5 \sV_1^2}{18}, \nonumber\\
\delta_1 & \approx & \frac{\U}{2} - \sum_{n=0}^{\infty} \left(\frac{1}{3}\right)^n  \sV_{n+1}  - \frac{2}{3}\U^2 + \frac{4}{3} \U\sV_1 - \frac{\sV_1^2}{3} ,\nonumber\\
\delta_2 &\approx& \sum_{n=0}^{\infty} \left(\frac{1}{3}\right)^n  \sV_{n+2}+ \U^2 -\frac{5}{2} \U \sV_1 + \sV_1^2,
\nonumber\\
\delta_p &\approx&(-1)^p \sum_{n=0}^{\infty} \left(\frac{1}{3}\right)^n  \sV_{n+p}, \qquad {(p>2)}
\end{eqnarray}
where we have performed one more iteration of $\epsilon_H$ to obtain $\epsilon_H/\U$ to ${\cal O}(N^{-2})$
since it appears
in the denominator of $\Delta_\curv^2$.

As noted above,
an advantage of the potential slow-roll parameters is that in the Taylor expansion for
$G'$ mixed $p$ terms in $\delta_p$ go away when rewritten in $\sV_p$,
 \begin{eqnarray}
G &\approx & \ln  \left( \frac{V}{12\pi^2\U} \right) -\frac{7}{6}\U -\frac{21}{8}\U^2 + \frac{7}{3}\U \sV_1 -\frac{1}{9}\sV_1^2, \nonumber\\
G' &\approx&  3\U - 2\sV_1 + \frac{17}{6} \U^2 -\frac{5}{3} \U \sV_1 -\frac{2}{3} \sV_1^2,\nonumber\\
G'' & \approx &-2 \sV_2   -6\U^2 + 8 \U \sV_1 , \nonumber\\
G^{(p)} & \approx & -2 \sV_p ,\qquad (p>2),
\label{eqn:GV}
\end{eqnarray}
for the scalars and
\begin{eqnarray}
G_h &\approx & \ln  \left( \frac{V}{6\pi^2} \right) -\frac{7}{6}\U -\frac{55}{24}\U^2 + \frac{17}{9}\U \sV_1, \nonumber\\
G'_h &\approx&  \U  + \frac{5}{2} \U^2 -2 \U \sV_1, \nonumber\\
G''_h & \approx &  -2\U^2 + 2 \U \sV_1 , \nonumber\\
G^{(p)}_h & \approx & 0 ,\qquad (p>2),
\label{eqn:GVt}
\end{eqnarray}
for the tensors.

With these relations, the full ${\cal O}(N^{-2})$ expressions with $\Delta N \ll 1$ for the curvature power spectrum observables,  
\begin{eqnarray}
\ln \Delta_\curv^2(k) &\approx& G(\ln x_f)  +  \sum_{p=1}^\infty \tq_p G^{(p)}(\ln x_f)  \nonumber\\ 
&& + \frac{\pi^2}{8} [G'(\ln x_f)]^2 -4 \left[\frac{f'}{f}(\ln x_f)\right]^2 , \nonumber\\
n_s-1 &\approx &  - G'(\ln x_f) -  \sum_{p=1}^\infty \tq_p G^{(p+1)}(\ln x_f) ,\nonumber\\  
\alpha &\approx & G''(\ln x_f) + \sum_{p=1}^\infty \tq_p G^{(p+2)}(\ln x_f),
\label{eqn:GseriesGen}
\end{eqnarray}
become 
\begin{eqnarray}
\Delta^2_\curv & \approx & \frac{V}{12\pi^2\U} \Big[ 1 +  
\left( 3\tq_1 -\frac{7}{6} \right) \U  - 2 \sum_{p=1}^\infty \tq_p \sV_p
\nonumber\\ 
&& + \left( 3 \tq_2 - \frac{2}{3} \tq_1 - \frac{103}{9} + \frac{3\pi^2}{2} \right) \U^2 
\nonumber\\
&& + \left( -4 \tq_2 + \frac{2}{3} \tq_1 + 15 - 2 \pi^2 \right) \U \sV_1
\nonumber\\
&& + \left( 4 \tq_2 - \frac{2}{3} \tq_1 - \frac{13}{3} + \frac{2\pi^2}{3} \right) \sV_1^2 \Big] ,
\nonumber\\
n_s & \approx & 1 - 3\U + 2\sV_1 + 2 \sum_{p=1}^\infty \tq_p \sV_{p+1} 
\nonumber\\ 
&& + \left( 6 \tq_1 - \frac{17}{6} \right) \U^2  -\left( 8 \tq_1 - \frac{5}{3} \right) \U \sV_1+ \frac{2}{3}  \sV_1^2  , 
\nonumber\\
\alpha & \approx & -2 \sV_2 - 2 \sum_{p=1}^\infty \tq_p \sV_{p+2} - 6\U^2 + 8 \U \sV_1 .
\label{eqn:SO}
\end{eqnarray}
The analogous relations for the tensor observables give
\begin{eqnarray}
\Delta^2_{+,\times} & \approx & \frac{V}{ 6\pi^2} \Bigg[ 1+
\left( \tq_1 - \frac{7}{6} \right) \U 
\nonumber\\
&& +\left( - \tq_2 + \frac{4}{3} \tq_1 - \frac{8}{3} + \frac{\pi^2}{6} \right) \U^2
\nonumber\\
&& + \left( 2 \tq_2 - 2 \tq_1 + \frac{17}{9} \right) \U \sV_1 \Bigg] , 
\nonumber\\
r & \approx & 8\U -16 \tq_1 \U(\U -\sV_1) ,  \nonumber\\
n_t & \approx & -\U + \left( 2 \tq_1 - \frac{5}{2} \right) \U^2 - 2 ( \tq_1 -1 ) \U \sV_1 ,
\nonumber\\
\alpha_t & \approx & -2\U(\U - \sV_1).
\label{eqn:SOt}
\end{eqnarray}
These expressions can be used to derive any second-order approximation
specified consistently by the order in the hierarchy of $\tq_p(\ln x_f)$ corrections or
in the standard approach by keeping only ${\cal O}(1/N\Delta N)$ terms with $\ln x_f=0$.

\begin{figure}[t]
\centering
\includegraphics[width=3in]{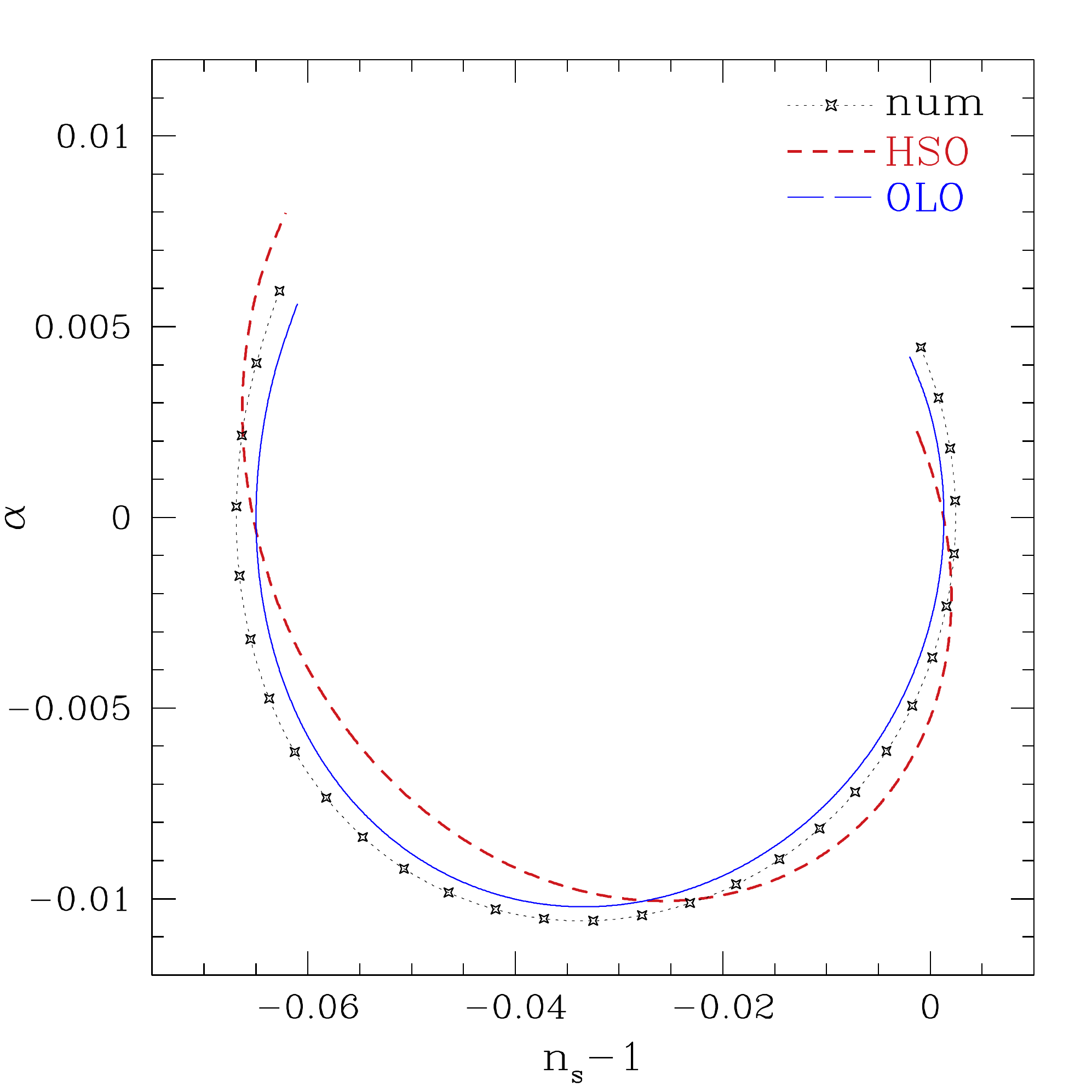}
\caption{Low frequency $\omega=1/3$ oscillation trajectories in the
  $n_s$-$\alpha$ plane under the OLO and HSO 
  approximations as in Fig.~\ref{fig:DN3Snsalpha}.
  HSO like SO provides an inconsistent trajectory and is worse than the simpler OLO approximation. }
\label{fig:DN3nsalphah}
\end{figure}

For example, in terms of the Hubble flow parameters of Eq.~(\ref{eqn:Hubbleepsp}), keeping 
the conversions in the standard approach where we assume $\epsilon_n = {\cal O}(1/N)$ gives
\begin{eqnarray}
n_s -1 &=& -2 \epsilon_1 -\epsilon_2 - 2\epsilon_1^2  + \left(2 \tq_1 - \frac{11}{3}\right) \epsilon_1\epsilon_2 \nonumber\\
&&+ \left(\tq_1 -\frac{1}{3} \right) \epsilon_2\epsilon_3, \nonumber\\
\alpha &=& -2 \epsilon_1 \epsilon_2 - \epsilon_2\epsilon_3 ,\nonumber\\
n_t &=& -2\epsilon_1 - 2\epsilon_1^2 + \left( 2 \tq_1 -\frac{8}{3}\right) \epsilon_1\epsilon_2 ,\nonumber\\
\alpha_t &=& -2\epsilon_1\epsilon_2, \qquad {\rm (HSO)} \label{eqn:HSO}
\end{eqnarray}
which reproduces the standard result used in the literature 
(e.g.~\cite{Planck:2013jfk,Ade:2015lrj,Baumann:2014cja}) with
$\tq_1=\tq_1(0) = 7/3 -\ln 2 -\gamma_E$.
Note that because the truncation is inconsistent in the models we consider with
$\Delta N \ll |N|$, these relations differ numerically from 
the SO approximation expressed in terms of the same truncation in the potential parameters 
in the text unless the potential is cubic.  
Nonetheless $n_s-1$ and $\alpha$ are still inconsistently evaluated, 
and the Hubble flow second order (HSO) also performs worse than the simpler OLO approximation 
(see Fig.~\ref{fig:DN3nsalphah}).

\eject


\bibliography{running}
\end{document}